\documentclass[10pt,a4paper]{article}
\usepackage{inputenc}
\usepackage{setspace}
\usepackage{amsmath}
\usepackage{amsfonts}
\usepackage{amssymb}
\usepackage{amsthm}
\usepackage{bm}
\usepackage{graphicx}
\usepackage{enumitem}
\graphicspath{ {./figures/} }
\newtheorem{cond}{Condition S\ignorespaces}

\newtheorem{remarkS}{Remark S\ignorespaces}
\usepackage[margin=1in]{geometry}
\usepackage[title]{appendix}
\usepackage{authblk}
\usepackage{xcolor}
\usepackage{times}
\usepackage{bm}
\usepackage{natbib}
\usepackage{graphicx}
\usepackage{float}
\usepackage{xcolor}
\usepackage{amsfonts}
\usepackage{dsfont}
\usepackage{bbm}
\usepackage{booktabs} 
\usepackage{caption}
\usepackage{subcaption}
\usepackage{mathrsfs} 
\usepackage{accents}
\usepackage{xr}
\usepackage{hyperref}
\usepackage{optidef}
\usepackage[mathscr]{euscript}
\usepackage{bbm}
\usepackage[algo2e]{algorithm2e} 
\newcommand{\bx}{\mathbf{x}}
\newcommand{\bX}{\mathbf{X}}
\newcommand{\bZ}{\mathbf{Z}}
\newcommand{\bz}{\mathbf{z}}

\usepackage{enumitem}
\usepackage{mathrsfs} 
\usepackage{accents}
\usepackage{algorithm}
\usepackage{xr}
\usepackage{adjustbox}
\definecolor{change}{rgb}{0,0.4,0}
\definecolor{ppr}{rgb}{0.129, 0.4, 0.675}
\makeatletter
\newcommand*{\addFileDependency}[1]{
  \typeout{(#1)}
  \@addtofilelist{#1}
  \IfFileExists{#1}{}{\typeout{No file #1.}}
}

\renewcommand{\algocf@captiontext}[2]{#1\algocf@typo. \AlCapFnt{}#2} 
\def\@algocf@capt@plain{top}
\renewcommand{\algocf@makecaption}[2]{%
  \addtolength{\hsize}{\algomargin}%
  \sbox\@tempboxa{\algocf@captiontext{#1}{#2}}%
  \ifdim\wd\@tempboxa >\hsize
    \hskip .5\algomargin%
    \parbox[t]{\hsize}{\algocf@captiontext{#1}{#2}}
  \else%
    \global\@minipagefalse%
    \hbox to\hsize{\box\@tempboxa}
  \fi%
  \addtolength{\hsize}{-\algomargin}%
}
\makeatother

\newcommand*{\myexternaldocument}[1]{%
    \externaldocument{#1}%
    \addFileDependency{#1.tex}%
    \addFileDependency{#1.aux}%
}

\myexternaldocument{supp_biometrika}

\newtheorem{theorem}{Theorem}
\newtheorem{remark}{Remark}

\usepackage{caption}
\usepackage[noend]{algpseudocode}

\usepackage{tikz}
\usetikzlibrary{positioning}

\tikzset{
    every neuron/.style={
        circle,
        draw,
        minimum size=1cm
    },
    neuron missing/.style={
        draw=none, 
        scale=4,
        text height=0.333cm,
        execute at begin node=\color{black}$\vdots$
    },
}

\usepackage{natbib}
\title{Efficient collaborative learning of the average treatment effect}
\author[*]{Sijia Li}
\author[*]{Rui Duan}
\affil[*]{Department of Biostatistics, Harvard T.H. Chan School of Public Health}

\date{}

\begin{document}
\maketitle

\begin{abstract}
In response to the growing need for generating real-world evidence from multi-site collaborative studies, we introduce an efficient collaborative learning approach to evaluate average treatment effect (ECO-ATE) in a multi-site setting under data sharing constraints. Specifically, ECO-ATE operates in a federated manner, using individual-level data from a user-defined target population and summary statistics from other source populations, to construct efficient estimator for the average treatment effect on the target population of interest. Our federated approach does not require iterative communications between sites, making it particularly suitable for research consortia with limited resources for developing automated data-sharing infrastructures. Compared to existing work data integration methods in causal inference,  ECO-ATE allows distributional shifts in outcomes, treatments and baseline covariates distributions, and achieves semiparametric efficiency bound under appropriate conditions. We conduct simulation studies to demonstrate the extent of efficiency gains achieved by incorporating additional data sources, as well as the robustness of our approach against varying levels of distributional shifts and overparameterization, compared to existing benchmarks. We apply ECO-ATE to a case study examining the effect of insulin vs. non-insulin treatments on  heart failure for patients with type II diabetes using electronic health record data collected from the \textit{All of Us} program. 
\end{abstract}

\maketitle

\section{Introduction}
\label{sec:intro}
With the increasing number of data networks and research consortia, there is growing interest in developing statistically and communication-efficient data fusion techniques to estimate causal effects across diverse focus areas { \citep{suchard2019comprehensive}. While the use of multiple data sources enables researchers to answer scientific questions with greater statistical power and broader generalizability, it is crucial to recognize the differences in treatment protocols, patient demographics, and data collection processes across healthcare systems, which can lead to substantial heterogeneity. }

A significant body of work has focused on addressing the challenges posed by distributional shifts. 
However, a key assumption underlying much of this research is the exchangeability condition, which assumes that a common conditional distribution of the outcome of interest is shared among these heterogeneous data sources \citep{rudolph2017robust}. Since the level of heterogeneity is only restricted to non-outcome variables, it is feasible to fuse data for estimating a causal effect in an unbiased and efficient way \citep{li2023efficient}.  However, such exchangeability condition may not hold in practice, especially when the set of covariates fails to fully capture the variability of the outcome among data sources. While some work offers relaxed exchangeability conditions \citep{guo2022multi,lee2023improving}, they still require certain distributional characteristics to be identical across populations, which does not fully resolve the challenges previously mentioned. 
Recently,  
\cite{li2025data} defined weakly aligned sources in which the ratio of conditional outcome distributions between these sources and the target distribution can be characterized by selection bias models, and thus accommodates a richer class of shape-constraints besides the ones imposed on the outcome mean functions.  Another line of work use  data-driven ways to determine whether to borrow from other data sources.   \cite{yang2023elastic} developed a test--and--pool procedure using a preliminary test statistics to first determine whether exchangeability holds. However, these adaptive data integration methods result in irregular estimators, making uniform inference challenging and performs poorly in small samples for certain data-generating process. In addition, many of the existing methods only benefit when transportability is likely to hold. When the exchangeability condition fails, these methods would introduce bias or loss of efficiency, which is also known as ``negative transfer".  

 Adding to these challenges, in  multi-site collaboration, individual-level data oftentimes cannot be shared across sites due to privacy concerns. { One common motivating scenario arises when working with multi-site electronic health record (EHR) data, where patient-level information cannot be shared across institutions due to health network policies. In such settings, federated learning methods allow researchers to conduct collaborative analyses without pooling individual observations \citep{brisimi2018federated}.} Therefore it is crucial to enable collaborative analysis in a federated way; namely, constructing estimators with access to only source-specific summary statistics. While many existing federated learning literature focuses on regression and classification settings \citep{li2022transfer}, few has focused on federated causal inference. \cite{xiong2023federated} and \cite{vo2022bayesian} defined their causal target estimand of interest on a combined population and assumes exchangeability. \cite{han2025federated} proposed a parametric federated adaptive estimator of the average treatment effect. However, their estimator is not efficient.

{ An illustrative real-world scenario motivating our approach involves studying how different diabetes treatments—particularly insulin and newer classes of glucose-lowering agents—impact heart failure (HF) outcomes across heterogeneous healthcare settings. Large multi-center studies have shown that sodium–glucose cotransporter-2 (SGLT2) inhibitors can confer a lower risk of HF hospitalization than other glucose-lowering medications \citep{kosiborod2017lower}, but the magnitude of this benefit appears to differ among various geographic and institutional contexts.   When data are scattered across different hospitals or regions—some of which may have limited HF endpoints—a privacy-preserving, decentralized data fusion approach is crucial to accurately estimate treatment effects while accounting for site-level heterogeneity.}

In this work, we propose a method for Efficient Collaborative learning of the Average Treatment Effect (ECO-ATE). We address the aforementioned challenges by allowing source-specific heterogeneity in the conditional outcome distributions, in additional to the ones in treatment mechanisms and covariates, between a user-specified target population and other sources. { 
        We propose a decentralized approach that uses individual-level data from the target population and summary-level statistics from other sources, achieving semiparametric efficiency under appropriate conditions. To achieve this, we propose tailored estimation strategies for nuisance parameters that respect privacy constraints, establish criteria for transferring summary-level information between sites, and design an efficient algorithm that minimizes communication costs. We identify the conditions under which ECO-ATE achieves the same asymptotic behavior as the pooled estimator, bridging the gap between theoretical efficiency and practical implementation. Additionally, we investigated comprehensively on the empirical performance of the proposed estimator constructed by different data-adaptive methods in practical settings and offer practical guidance on implementation, in the hope of making federated data fusion an accessible and reliable tool. }

\section{Problem Setup}

\label{sec:notations}
We use uppercase letters to denote random variables and lowercase letters for their realizations. When uppercase letters represent distributions, the corresponding lowercase letters denote their density functions.  We condition on lowercase letters in expectations to indicate conditioning on a random variable taking a specific value. We use $[k]$ to denote $\{1,\ldots,k\}$.  We let $E_{P}$ denote the expectation operator under a distribution $P$, and {let $\mathds{P}_{n}$ denote the empirical measure such that  $\mathds{P}_{n}O:= \frac{1}{n}\sum_{i=1}^{n}(O_i)$. } 
{For a list of vectors $v_l$, we write $(v_l)_{l\in\mathcal{L}}$ to denote the concatenation of these vectors.} {We use $R_{ab}$ to denote the element in the $a^\mathrm{th}$ row and $b^\mathrm{th}$ column of the matrix $R$. }

Our goal is to estimate the average treatment effect in a target population $Q^0 \in \mathcal{Q}$, where $\mathcal{Q}$ is nonparametric. We let $\bX$ denote $d$-dimensional baseline covariates, $A$ denote the indicator of being treated and $Y$ denote the outcome of interest. Under positivity, consistency and no unmeasured confounding assumptions \citep{rubin1980randomization,rosenbaum1983central}, the target average treatment effect can be identified as $\psi(Q^0) =E_{Q^0}\left[E_{Q^0}[Y\mid A=1,\bX]\right] - E_{Q^0}\left[E_{Q^0}[Y\mid A=0,\bX]\right].$

Suppose we have access to individual-level data of the target population collected from a target site. In addition to the target site, there are $k$ source sites in which we observe the same data structure, but the underlying population may be different to the target population. We let $S \in \mathcal{S}:= \{0 \cup [k]\}$ denote the site indicator, where $S = 0$ indicate the target site, and $S\in[k]$ each indicates a source site. Together, we {observe $n$ i.i.d copies of } $(\bZ,S) = (\bX,A,Y,S) \sim P^0 \in \mathcal{P}$. 
Since $P^0(\cdot \mid S= 0 ) = Q^0$, the target average treatment effect can be always identified using the target site data. For clarity, we denote the identified parameter as a functional of $P^0$ such that $\phi(P^0) = \psi(Q^0)$ where
$\phi(P^0) =E_{P^0}\left[E_{P^0}[Y\mid A=1,\bX,S=0] \mid S=0\right] - E_{P^0}\left[E_{P^0}[Y\mid A=0,\bX,S=0] \mid S=0\right].$

Despite the distributional shifts among sites, using source site's information may still be helpful for estimating the target estimand $\phi(P^0)$. For the distributional shifts of the covariate and treatment assignment, we assume that for any source site $s\in [k]$,  $p^0(\bx,a \mid s)$ can be different  from the target distribution $p^0(\bx,a \mid S=0)$ without knowing how they are different, although the source distributions need to have sufficient overlap with the target distribution, which will be further discussed in Section~\ref{sec:app:regularity}. Existing data fusion work often assumes an exchangeability on the conditional outcome distribution across populations, that is, for each $s \in [k]$:
\begin{align*}
    p^0(y\mid a,\bx, s) = p^0(y\mid a,\bx,S=0). 
\end{align*}
As a result, the distributional shifts 
across sites can be handled by reweighing the source data points properly by the ratio of
$p^0(a,\bx\mid S=0)/p^0(a,\bx \mid s)$, given that there are sufficient overlap between the two distributions. 
In this work, we consider a more challenging scenario where due to {unmeasured site-level effect modifiers},
exchangability is likely violated. Instead, we allow shifts in the conditional distributions, and propose to model such shift by a flexible semiparametric density ratio model. For each site $s \in [k]$, we assume that
\begin{align*}
    p^0(y\mid a,\bx, s) = w_s^*(\bz;\beta^0_s,W_s)p^0(y\mid a,\bx,S=0), 
\end{align*}
with $w_s^*(\bz;\beta^0_s,W_s):= w_s(\bz;\beta^0_s)/W_s(\bx,a;\beta^0_s)$ and $W_s(x,a;\beta^0_s) := E_{P^0}[w_s(\bZ;\beta^0_s)\mid \bX=\bx,A=a, S =0]$, where the form of the site-specific weight function $w_s(z;\beta^0_s)$ is known, and the parameters associated with the model, $\beta^0_s$ is unknown. 
In other words, the shift in the conditional outcome distributions between target and source sites are known up to a finite-dimensional parameter $\beta^0: = (\beta^0_s)_{s\in [k]} \in\mathcal{B}$. {When properly characterizing the misalignment between source and target data, this framework benefits from source datasets that were previously excluded by existing data fusion approaches,  enabling the calibration of such sources to unlock further efficiency gains. }

\section{Efficient federated learning algorithm}
\label{sec:methods}
\subsection{Overview}
\label{overall step}
We first provide an overview of the main steps of the proposed ECO-ATE method. We will use knowledge from semiparametric efficiency theory to derive the canonical gradient of the average treatment effect under the proposed framework, and develop a federated inferential method where summary-level information of the canonical gradient is shared across sites to account for site-level heterogeneity. The procedure begins with the target site estimating distributional shifts for each source site using summary statistics collected from those sites. Following this, nuisance estimates are broadcast to all sites, serving as foundational elements for constructing the site-specific canonical gradient. Next, each source site evaluates the canonical gradient  and sends these summaries back to the target site. The target site then assembles the ECO-ATE estimator using the collected summaries from all source sites. During this procedure, each site participates in only two rounds of communication, making the process communication-efficient and easy to implement in practice. The detailed steps are summarized in Algorithm \ref{alg:ate}.

\subsection{Target site estimates distributional shifts}
\label{step1}
The first step is to characterize the degrees of distributional shifts for each source site using target data and summary statistics from the source sites. The key of our method is to correctly adjust for the distributional shifts between $p(\cdot \mid S=s)$ and $p(\cdot \mid S=0)$ for each source site $s \in [k]$. For estimating the target average treatment effect, it is natural to divide the distributional shifts into two layers. One involves shifts in the covariates and treatments, where we denote $\lambda_s(\bx,a):= p^0(\bx,a\mid S=s)/p^0(\bx,a\mid S=0)$ as the density ratio of the covariates and treatment mechanism between source site $s$ and the target, and denote $\lambda(\bx,a):= (\lambda_s(\bx,a))_{s\in[k]}$.  The other involves the shift in  the conditional outcome distribution $Y\mid A, \bX$. Each site is required to specify its site-specific form of weight function for the conditional outcome distribution shift, i.e.,$w_s(\bz;\beta^0_s)$. An example of such a function would be  exponential tilt density ratio model, in which we specify 
\[
w_s(\bz;\beta^0_s) = \exp\{{\beta^0_s}^\top \xi_{s}(y, a, \bx)\}
\]
where $\xi_{s}$ are prespecified basis functions. When we have centralized data from all sites, {$\beta^0_s$ can be estimated via maximum likelihood using an estimate for the normalizing function $E_{P^0}[w_s(\bZ;\beta^0_s)\mid \bX,A,S=0]$, of which can be obtained using kernel regressions \citep{nadaraya1964estimating}, or other nonparametric data-adaptive approaches.} 
In a federated setting, since only aggregated information is allowed to be shared across sites, we propose to estimate the density ratio models via the method of moments. The underlying intuition is that, by correctly adjusting for the distributional shifts, we will obtain sufficient (in fact, infinite) moments to be matched. Consequently, an initial estimator of $\beta^0_s$ can be constructed by solving the following estimating equation in the target site: 
\begin{align}
    \bar{\xi}_{s}  &=  \frac{1}{n_0}\sum_{i \in \mathrm{target}} \hat{\lambda}_s(\bx_i,a_i)w^*_s(\bz_i;\beta^0_s,\hat{W}_s)\xi_{s}(y_i, a_i,\bx_i) \label{MoM1}
\end{align}
where $\bar{\xi}_{s} = \mathds{P}_{n,s}\xi_{s}(\bZ_i)$ is the empirical mean of $\xi_{s}(\bZ)$ calculated and shared by the $s$-th source site. 
The estimated density ratio $ \hat{\lambda}_s(\bx,a)$ and the normalizing function $\hat{W}_s(\bx,a;\beta)$ for any $\beta \in \mathcal{B}$ {can both be estimated via 
any applicable data-adaptive methods including exponential tilting models \citep{efron1978geometry}, generalized additive models \citep{hastie2017generalized} and methods of sieves \citep{grenander1981abstract}. Alternatively, each source can estimate its own $\hat{p}(\bx,a\mid S=s)$ via methods such as wavelets density estimation \citep{donoho1996density}. The key is that these estimators are essentially functions of $(\bx,a)$, and  they need to be  evaluated in the target data using summary statistics without accessing the individual-level data from source sites. { Note that \eqref{MoM1} uses the density ratio $\hat{\lambda}_s$, which calibrates the density of $(\mathbf{X},A)$ for data source $s$ to the one of the target. Although, by a change of measure and under overlap, any source \(m\in[k]\) could in principle serve as the anchor, we anchor at the target for two reasons. First and most importantly, it enables a single-round federated implementation: sources transmit their moments once. By contrast, using a non-target anchor typically requires multiple communication rounds. Second, the target-anchored weight appears in the canonical gradient as shown in Theorem~\ref{thm:canonical}. This choice eliminates the need to estimate additional density ratios, calibrates all sources directly to the target for clearer interpretation, and typically improves stability. } 

{While we highlight the flexibility of approaches that can be employed under the proposed framework, we want to emphasize that various approaches may lead to different performances for the resulting estimator depending on whether the required regularity conditions in Section~\ref{sec:app:regularity} are satisfied. We also acknowledge that as the dimension of $\bX$ increases, the choice of estimation strategies for the nuisance functions may have a more pronounced effect, and additional computational challenges may arise that need to be considered.}  To emphasize on the federated nature, we denote these estimates as $\hat{\lambda}_s(\bx,a; \hat{\gamma}_s)$ and $\hat{W}_s(\bx,a;\hat{\beta}_s, \hat{\zeta}_s)$, where $\hat\gamma_s$ and $\hat\zeta_s$ denote summary statistics.} If these estimators are consistent, the method of moment estimate $\hat{\beta}_s$ is consistent.

\subsection{Target site broadcasts to all source sites}
\sloppy To prepare for the construction of an efficient estimator for $\phi(P^0)$, we require the target sites to broadcast a list of summary statistics.
For ease of reading and clarity, we begin by introducing some notation. 
For a fixed $s$, let $\dot{w}_{s}$ be the derivative of $w_s$ with respect to $\beta_{s}$ evaluated at $\beta^0_{s}$. Let $r_{s}(\bz;\beta,W,\lambda) :=  r(\bz;\beta,W,\lambda)P^0(S=s)\lambda_s(\bx,a)w^*_{s}(\bz;\beta_{s},W_s)$ with $r(\bz;\beta,W,\lambda) := \left\{\sum_{s \in \mathcal{S}} w^*_{s}(\bz;\beta_{s},W_s)\lambda_s(\bx,a)P^0(S=s) \right\}^{-1}$. In addition, we let $\bar{w}^*:= (w^*_{s})_{s \in \mathcal{S}}$ and $\Delta$ be the diagonal matrix with diagonal $(P^0(S=s)_{s \in \mathcal{S}})^\top$. We define an $(k+1) \times (k+1) $ matrix $M(\bx,a;\beta,W,\lambda) = \Delta^{-1} - \int r(\bz;\beta,W,\lambda)\bar{w}^*(\bz;\beta,W){\bar{w}}^{*\top}(\bz;\beta,W)\,P^0(dy\mid a,\bx, S=0) $ and let $M^{-}$ be the generalized inverse of $M$. 
We let $\tilde{a}(\bz;\beta,W,\lambda,P^0) :=\sum_{m \in \mathcal{S}} r_{m}(\bz;\beta,W,\lambda)\dot{\ell}_{\beta_{s}}(\bz,m;\beta_{m},P^0)$, where $    \dot{\ell}_{\beta_{s}}(\bz,s';\beta_{s},P^0) := \frac{\dot{w}_{s}(\bz,s';\beta^0_{s})}{w_{s}(\bz,s';\beta_{s})} -E_{P^0}\left[ \frac{\dot{w}_{s}(\bZ,S;\beta^0_{s})}{w_{s}(\bZ,S;\beta_{s})}\mid  A=a,\bX=\bx, S=s'\right]$ is the score function of $\beta^0_s$ relative to the model where $Q^0$ is known. 

Specifically, the target site will broadcast estimators of the following:
\begin{enumerate}[label=(\alph*),leftmargin=*]
    \item \sloppy Nuisance parameters that measure distributional shifts of all source sites: sample size of each source site, $\beta^0$, $\lambda(\bX,A;{\gamma})$, form of the basis functions $\xi:= (\xi_s)_{s\in [k]}$, normalizing functions $ {W}(\bX,A;{\beta^0})$, $E_{P^0}[r(\bZ;\beta^0,{W},{\lambda}) \bar{w}^*(\bZ;\beta^0,{W})\mid A,\bX, S=0]$, and $E_{P^0}[r(\bZ;\beta^0,{W},{\lambda}) \bar{w}^*(\bZ;\beta^0,{W}){\bar{w}^*}^\top(\bZ;\beta^0,{W})\mid A,\bX, S=0]$.
    \item \sloppy Nuisance parameters for the target average treatment effect: {${\pi}(A,\bX):= P^0(A\mid \bX, S=0)$, ${\mu}(A,\bX):= E_{P^0}[Y\mid A,\bX, S=0]$,} $E_{P^0}[\tilde{d}(\bZ;\beta^0,{W},{\lambda},P^0)\mid A,\bX, S=0]$ and $E_{P^0}[\tilde{d}(\bZ;\beta^0,{W},{\lambda},P^0) \bar{w}^*(\bZ;\beta^0,{W})\mid A,\bX, S=0]$, where $\tilde{d}(\bZ;\beta^0,{W},{\lambda},P^0) :=  r(\bZ;\beta^0,{W},{\lambda})\sum_{a=0}^1\frac{2a - 1}{{\pi}(a,\bX)} \left(Y - {\mu}(a,\bX) \right)$.
    \item \sloppy Nuisance parameters for estimating $\beta^0$: $E_{P^0}[\tilde{a}(\bZ;\beta^0,{W},{\lambda},P^0)\mid A,\bX, S=0]$, and $E_{P^0}[\tilde{a}(\bZ;\beta^0,{W},{\lambda},P^0) \bar{w}^*(\bZ;\beta^0,{W})\mid A,\bX, S=0]$.
\end{enumerate}

{In the above, conditional expectations can be estimated using different aforementioned data-adaptive approaches such as exponential tilting models \citep{efron1978geometry}, generalized additive models \citep{hastie2017generalized} and methods of sieves \citep{grenander1981abstract}, such that a set of summary-level statistics can be shared across sites to evaluate these conditional expectations at a given site.

Instead of defining the summary statistics for each conditional mean,  we collectively define $\hat\theta$ as the list of summary statistics needed for estimating all these conditional expectations. Accordingly, we slightly abuse the notation and use $E_{\hat{P}_\theta}$ to denote the estimated conditional expectations.} After the broadcast, each source site not only obtains its site-specific nuisance estimates but also the ones for all other sites. It is important to note that the knowledge about distributional shifts in other source sites is crucial for efficiently estimating $\beta^0_s$ and therefore $\phi(P^0)$. This is because all sites are intertwined via the target population -- knowing about others shifts inadvertently informs the underlying $P^0(\cdot \mid S=0)$. 

\vspace{-1 cm}

\subsection{Transfer site-specific knowledge and efficient fusion}
\label{step2}

{Using the broadcast nuisance parameters in categories (a) and (b)},
each site $s\in\mathcal{S}$ constructs and sends to target the following site-specific  summary:
\begin{align*}
    \mathcal{H}_s & =\mathds{P}_{n,s} \left(\mathcal{L}(\tilde{d})(\bZ_i;\hat{\beta},\hat{P}_{\theta}) - E_{\hat{P}}[\mathcal{L}(\tilde{d})(\bZ_i;\hat{\beta},\hat{P}_{\theta}) \mid A_i, \bX_i, S_i]\right),
\end{align*}
where $\mathds{P}_{n,s}O$  denotes the empirical mean  over subjects with $S_i=s$, and we define the linear operator $\mathcal{L}(\tilde{d})(\bZ;\beta^0,P^0):=\tilde{d}(\bZ;\beta^0,W,\lambda,P^0)- E_{P^0}\left[\tilde{d}(\bZ;\beta^0,W,\lambda,P^0)\mid  A,\bX,S=0\right] + E_{P^0}\big[\tilde{d}(\bZ;\beta^0,W,\lambda,P^0) {\bar{w}}^{*\top}(\bar{\bZ};\beta^0,W)\mid  A,\bX,S=0\big]  M^{-}(\bX,A;\beta^0,W,\lambda)^{\top} \bigg\{{\bar{w}}^{*}(\bZ;\beta^0,W)r(\bZ;\beta^0,W,\lambda)-E_{P^0}\left[{\bar{w}}^{*}(\bZ;\beta^0,W)r(\bZ;\beta^0,W,\lambda)\mid  A,\bX,S=0\right] \bigg\}$.  
Similarly, the term $\mathcal{L}(\tilde{d})(\bZ_i;\hat{\beta},\hat{P}_{\theta})$ represents the substitution of $\beta^0$, $W$, $\lambda$, and all other conditional expectations with their estimates (i.e., $E_{\hat{P}_\theta}$). The term $E_{\hat P}[d^*(\bZ; \hat{\beta}, \hat{W}, \hat{\lambda}, \hat{P}_{\theta})\mid A, \bX, S]$ represents an estimator for $E_{P^0}[d^*(\bZ; \beta^0, W, \lambda, P^0) \mid A, \bX, S]$ constructed within site $s$. We can use more flexible estimators, such as kernel regression, to estimate $E_{P^0}[d^*(\bZ; \beta^0, W, \lambda, P^0) \mid A, \bX, S]$,  not limiting to  methods that require evaluation on data from different sites using a set of summary statistics. Consequently, we use the notation $E_{\hat{P}}$, in contrast to the conditional mean estimators denoted as $E_{\hat{P}_\theta}$.

 We now construct the remaining piece to account for the estimation of $\beta^0$. It can be verified that the efficient score function of  $\beta^0_{s}$ takes the form of 
\begin{align*}
    \dot{\ell}^*_{\beta_{s}}(\bz,s';\beta^0,P^0) &= \dot{\ell}_{\beta_{s}}(\bz,s';\beta^0,P^0)- \left\{\mathcal{L}(\tilde{a})(\bz;\beta^0,P^0) - E_{P^0}\left[ \mathcal{L}(\tilde{a})(\bZ;\beta^0,P^0)\mid a,\bx, s'\right]\right\}.
\end{align*}
Plugging in nuisance estimates in categories (a) and (c) received from the target site, each source can construct the efficient score functions for all $\beta^0$ evaluated on its site-specific data. Specifically, each site $s \in \mathcal{S}$ will construct and send to target the following summaries:
\begin{align*}
    \mathcal{L}_s  &= \mathds{P}_{n,s}  \dot{\ell}^*(\bZ_i,S_i;\hat{\beta},\hat{W},\hat{\lambda},\hat{P}_{\theta});\quad
    \mathcal{I}_s = \mathds{P}_{n,s} \{\dot{\ell}^*(\bZ_i,S_i;\hat{\beta},\hat{W},\hat{\lambda},\hat{P}_{\theta}) {\dot{\ell}^*}^\top(\bZ_i,S_i;\hat{\beta},\hat{W},\hat{\lambda},\hat{P}_{\theta}) \}
\end{align*}
Finally, the target site will compute $\mathcal{I}:= \sum_{s \in \mathcal{S} }P^0(S=s)\mathcal{I}_s$, and construct quantities $(\mathcal{M}_s)_{s\in\mathcal{S}}$:
\begin{align*}
    \mathcal{M}_s  &=  E_{\hat{P}}\bigg[\bigg\{ \sum_{a=0}^1\frac{2a-1}{\hat{\pi}(a,x)} \left(Y - \hat{\mu}(a,\bx) \right) \bigg\}\dot{\ell}^*(\bZ,S;\hat{\beta},\hat{W},\hat{\lambda},\hat{P}_{\theta}) \mid S=0\bigg]{\mathcal{I}}^{-1}\mathcal{L}_s.
\end{align*}
Finally, our proposed ECO-ATE estimator takes the form of $$\hat{\phi}_{\textnormal{ECO-ATE}} =  \frac{1}{k+1}\sum_{s\in \mathcal{S} }(\mathcal{H}_s + \mathcal{M}_s) + \mathcal{N}_0 ,$$ where $\mathcal{N}_0 = \mathds{P}_{n,0}\left(\hat{\mu}(1,\bX_i)  -\hat{\mu}(0,\bX_i)\right)$.

\begin{algorithm}[H] \label{alg}
    \caption{Efficient collaborative learning of the average treatment effect}\label{alg:ate}
        \textbf{1. Target estimate distribution shifts:}
        \begin{enumerate}[label=\roman*.,leftmargin=*]
            \item Each \textbf{source site} send: sample size, $\hat{\gamma}_s$, form of $w_s$, and summary $\bar{\xi}_s$ to target site.
            \item \textbf{Target site} estimates shifts for each $s\in [k]$  by matching moments in \eqref{MoM1}. 
            \item \textbf{Target site} broadcast the following estimated nuisance parameters to all source sites: 
            \begin{enumerate}[label=(\alph*)]
                 \item \sloppy For measuring shifts: sample sizes, $\hat{\lambda}(\bx,a;\hat{\gamma})$,  forms of $w$, $\hat{\beta}$, {forms of all} $\xi$, $ \hat{W}(\bx,a;\hat{\beta})$, $E_{\hat{P}_\theta}[r(\bZ;\hat{\beta},\hat{W},\hat{\lambda}) \bar{w}^*(\bZ;\hat{\beta},\hat{W})\mid A,\bX, S=0]$, and $E_{\hat{P}_\theta}[r(\bZ;\hat{\beta},\hat{W},\hat{\lambda}) \bar{w}^*(\bZ;\hat{\beta},\hat{W}){\bar{w}^*}^\top(\bZ;\hat{\beta},\hat{W})\mid A,\bX, S=0]$.
                \item For estimating ATE:  $\hat{\pi}(A,\bX)$, $\hat{\mu}(A,\bX)$, $E_{\hat{P}_\theta}[\tilde{d}(\bZ;\hat{\beta},\hat{W},\hat{\lambda},\hat{P}_{\theta})\mid A,\bX, S=0]$ and $E_{\hat{P}_\theta}[\tilde{d}(\bZ;\hat{\beta},\hat{W},\hat{\lambda},\hat{P}_{\theta}) \bar{w}^*(\bZ;\hat{\beta},\hat{W})\mid A,\bX, S=0]$. 
                \item For estimating $\beta^0$: $E_{\hat{P}_\theta}[\tilde{a}(\bZ;\hat{\beta},\hat{W},\hat{\lambda},\hat{P}_\theta)\mid A,\bX, S=0]$ and $E_{\hat{P}_\theta}[\tilde{a}(\bZ;\hat{\beta},\hat{W},\hat{\lambda},\hat{P}_\theta) \bar{w}^*(\bZ;\hat{\beta},\hat{W})\mid A,\bX, S=0]$.
            \end{enumerate}
        \end{enumerate}
     \textbf{2. Transfer site-specific learnings and efficient fusion:}
     \begin{enumerate}[label=\roman*.,leftmargin=*]
         \item Each \textbf{source site} construct and send $\mathcal{L}_s$, $\mathcal{I}_s$ and $\mathcal{H}_s$,  to the target site.
         \item \textbf{Target site} construct $\mathcal{L}_0$, $\mathcal{I}_0$, $\mathcal{H}_0$, and quantities $(\mathcal{M}_s)_{s\in\mathcal{S}}$.The proposed ECO-ATE estimator is $$\hat{\phi}_{\textnormal{ECO-ATE}} =  \frac{1}{1+k}\sum_{s\in\mathcal{S}}(\mathcal{H}_s + \mathcal{M}_s) + \mathcal{N}_0 .$$ 
     \end{enumerate} 
\end{algorithm}

\section{Theoretical guarantees}
\label{sec:theory}
We now present our main theorem on the canonical gradient of the target average treatment effect.
\begin{theorem}\label{thm:canonical}
    Suppose each weight function $w_s$ is differentiable in $\beta_s$ at $\beta^0_s$. Under Condition S~\ref{cond:ide}, the canonical gradient of the target average treatment effect $\phi$ relative to $\mathcal{P}$ at $P^0$ is 
    \begin{align*}
        D^\mathrm{eff}_{P^0}(\bz,s;\beta^0) & = \mathcal{L}(d)(\bz;\beta^0,P^0) - E_{P^0}[\mathcal{L}(d)(\bZ;\beta^0,P^0) \mid a,\bx,s]  \\
        & \quad + \frac{\mathds{1}(s =0)}{P^0(S=0)}\left(\mu(1,\bx)  -\mu(0,\bx) - \phi(P^0)\right) \\
        & \quad + E_{P^0}\left[\left\{\sum_{a=0}^1 \frac{2a-1}{\pi(a,\bX)}(Y - \mu(a,\bX))\right\} \dot{\ell}^*(\bZ,S;\beta^0,P^0)\right]D^\beta_{P^0}(\bz,s;\beta^0), \label{eq:canonical}
    \end{align*}
    where $D^\beta_{P^0}(\bz,s;\beta^0)  = E_{P^0}[\dot{\ell}^*(\bZ,S;\beta^0,P^0){\dot{\ell}^*}^\top(\bZ,S;\beta^0,P^0)]^{-1}\dot{\ell}^*(\bz,s;\beta^0,P^0)$.
\end{theorem}

The algorithm outlined in Section~\ref{sec:methods} constructs each of these components step-by-step. Specifically, step 1 collects the necessary nuisance estimates; step 2 constructs pieces of the canonical gradient $D^\mathrm{eff}_{P^0}$. The algorithm finishes with constructing ECO-ATE in the form of a one-step estimator. In Section~\ref{sec:app:regularity}, we further state the regularity conditions under which the proposed ECO-ATE estimator achieves the semiparametric efficiency bound. 

To see how ECO-ATE handles heterogeneity and achieves efficiency gains, it is helpful to separate two layers of shift. First, we allow arbitrary heterogeneous but well-overlapping $(\mathbf{X},A)$ distributions across sources and correct via flexible density ratio estimation. Under Conditions S\ref{cond:ide} and S\ref{lem:conditions}, the rate at which the these shifts are estimated affects ECO-ATE only at second order. By contrast, first-order efficiency gain arises from the assumption that outcome shift can be characterized by a structured low-dimensional selection-bias models , i.e. the additional knowledge on the form of the parametric weight functions $w_s$ for each $s\in [k]$. When these forms are correctly specified, ECO-ATE attains a semiparametric efficiency bound that is no larger than variance of the target-only estimator. The gain diminishes as the dimension of $\beta$ grows, and vanishes if $w_s$ is fully nonparametric, since the cost of estimating $\beta$ eventually offsets any efficiency improvement.

\begin{remark} [Prevent negative transfer]
    {Under Conditions S\ref{cond:ide} and S\ref{lem:conditions},  ECO-ATE is guaranteed against negative transfer. Incorporating data from a source site will not lead to bias or loss of efficiency compared to the target-only estimator, regardless of the level of distributional shifts. This is a direct result of $D^{\mathrm{eff}}_{P^0}$ being the canonical gradient of the target average treatment effect. }
\end{remark}
{When implementing ECO-ATE in practice, each source site $s$ will have to make an educated guess on its form of shift $w_s$ relative to the target site. When the outcome is binary, $\beta^0$ would correspond to the shifts in log odds in different stratification schemes, which can be determined based on historical data and domain knowledge. Alternatively, source site can overparameterize $w_s$ by increasing the dimension of $\beta^0_s$. With overparameterization, there will be efficiency loss comparing to a correctly specified parsimonious model, but the efficiency of the ECO-ATE estimator will never be worse than the one not including the source site, providing a safeguard even there is a lack of prior domain knowledge. }

\section{Simulation}
\label{sec:simulation}
\sloppy We simulated one target site and three source sites, each with a fixed sample size of 500 observations. Conditional on data source $S$, we generated covariate $\bX \in \mathbb{R}^{10}$ and treatment A based on the following data generating mechanism: {  $\bX\mid S \sim \textnormal{N}(\mu_S,I)$, where $\mu_0 = (0,0,0.1,0,0.2,0.05,-0.3,0,0,0.1)$, $\mu_1 = (0,-0.05,0.1,-0.05,0.2,0.125,-0.3,0,0,0.1)$, $\mu_2 = (0,0.1,0.1,0.1,0.2,0.05,-0.3,0,0,0.1)$, $\mu_3 = (0,-0.1,0.1,-0.1,0.2,0.15,-0.3,0,0,0.1)$,} and $A\mid (\bX,S) \sim \textnormal{Bernoulli}(1/(1+\exp(0.05X_2)))$. We generate {$Y\mid (A,\bX,S) \sim \textnormal{Gamma}\{\alpha^\top (\bX + A\bX) + 10A + 10 - \epsilon \mathds{1}(S=1)X_6 -  \epsilon \mathds{1}(S=2)A-  \epsilon \mathds{1}(S=3)AX_6  , \alpha^\top \bX +10 \}$ where $\alpha = (1,-0.5,0,0,0,0.5,0,0,0,0)$ and $\epsilon = (0, 0.2, 0.5,0.7)$}, with $\epsilon=0$ indicates perfect alignment between data sources and large $\epsilon$ implies weaker alignment.  Under the current setup, each data source has a distinct form of weight function: {
$w_1(\bz;\beta_1^0) = \exp({\beta_{1}^0}x_6\log y)$, $w_2(\bz;\beta_2^0) = \exp(\beta^0_{2}a\log y) $, and $w_3(\bz;\beta_3^0) = \exp(\beta^0_{3}x_6a\log y)$.} Additionally, we examined a scenario where, instead of supplying the true weight functions, we estimated overparameterized weight functions. We provide more detailed description of the overparametrization scheme in Section~\ref{sec:app:overpar}. 

We estimate the average treatment effect and compare three types of estimators under varying extents of shifts in the conditional outcome distributions:  (1) a target-only estimator which only uses target data for estimation, (2) a na\"ive fusion estimator that assumes exchangebility in the conditional outcome distributions across all sources {($w_s=1$ for all $s\in[k]$)}, and (3) { the ECO-ATE estimators as outlined in Algorithm~\ref{alg:ate} using all sites but with different data-adaptive approaches for estimating nuisance parameters.}  Initial estimates of $\beta^0$ were obtained via method of moments as outlined in Step 1 in Algorithm~\ref{alg:ate}. { During this stage, different estimation approaches were compared, including linear regression and Lasso regression \citep{tibshirani1996regression}. When broadcasting $\hat{W}$, we compared different data-adaptive methods such as random forest, support vector machines, neural network and gradient boosting.} We used the exponential tilt model for modeling the density ratios of $\bX$ and $A\mid \bX$. { In addition, the conditional expectations involving $\hat{\beta}$ and $\hat{W}$ were estimated via simple linear regression or SuperLearner \citep{polley2010super} with a library consisting of linear regression and random forest for comparison. }   {We use the R package \texttt{glmnet} with tuning parameters selected via 10-fold cross-validation, and the R package \texttt{SuperLearner} where the weights for each algorithms are selected via 10-fold validation. }Propensity scores were estimated via main terms linear-logistic regression. 
500 Monte Carlo replications were conducted. 

\begin{figure}[H]
    \centering
    \includegraphics[width=\linewidth]{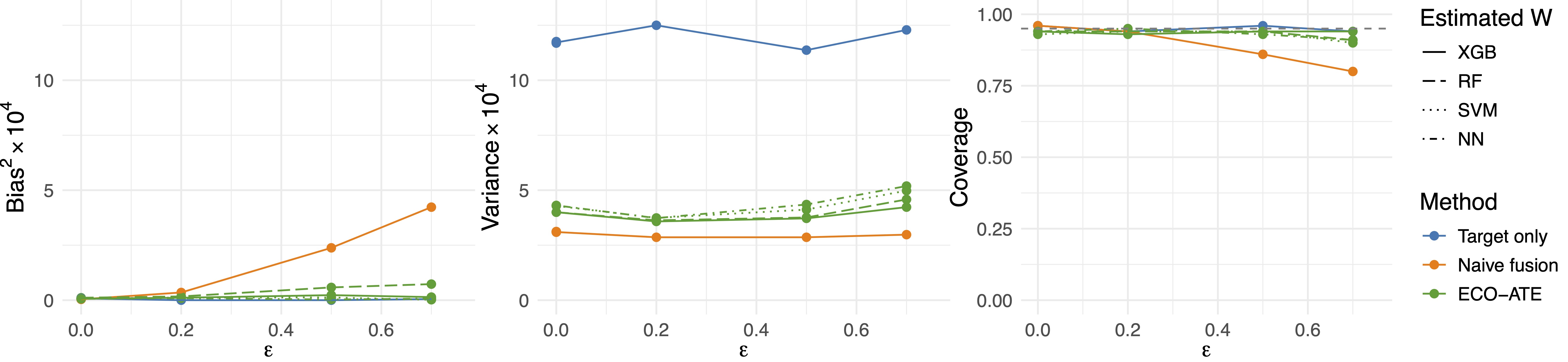}
        \caption{{Scaled bias squared, variance and coverage of the target only (blue), naive fusion (orange) and proposed ECO-ATE estimators (green). For ECO-ATE estimators, $\beta^0$ was estimated via linear regression while different models were employed for estimating $W$ as indicated by different line types.}}
    \label{fig:results_lane1}
\end{figure}

Figure~\ref{fig:results_lane1} displays the main results. 
The na\"ive fusion outperforms all estimators in the absence of shifts. This is expected, since ECO-ATE assumes weak alignment instead of full exchangeability and spends additional efforts in estimating $\beta^0$, leading to some loss of efficiency compared to na\"ive fusion. However, as the degree of alignment diminishes, na\"ive fusion is unable to distinguish such misalignment and leads to biased estimates. In contrast, ECO-ATE estimators are always consistent across varying degrees of alignment in both $Y$ and $\bX$, and have nominal coverage. { Various estimation approaches for $W$ gave similar performance when the extent of shifts is low; among which, gradient boosting achieves the smallest variance. When the weight functions are overparametrized, the efficiency gain is reduced as expected when overparametrization is moderate (Figure~\ref{fig:results_lane1}) with slight under-coverage, given the limited sample size.}

\begin{figure}[H]
    \centering
    \includegraphics[width=\linewidth]{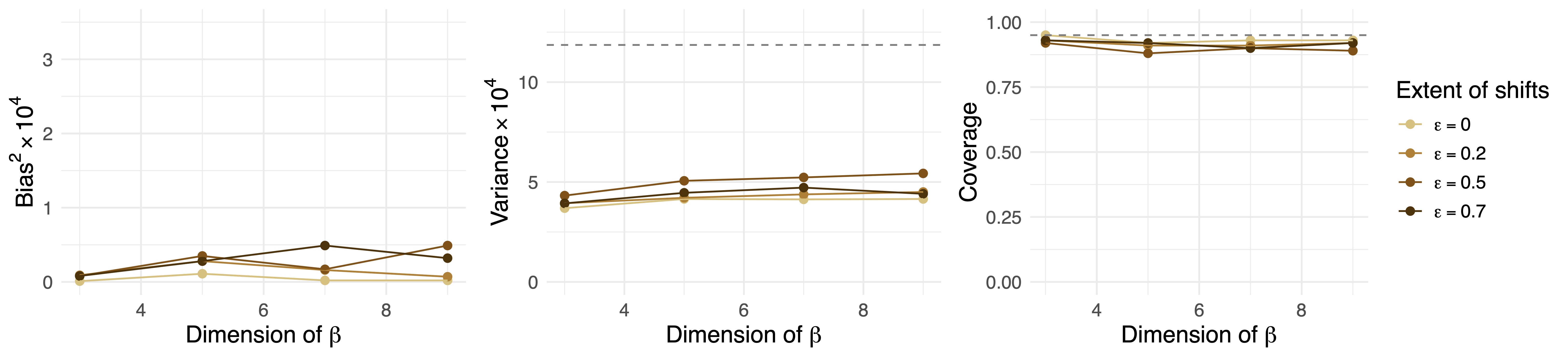}
    \caption{{Scaled bias squared, variance and coverage of the proposed ECO-ATE estimators across different overparameterization schemes. Throughout, $\beta^0$ was estimated via the LASSO and $W$ was estimated by gradient boosting.}}
    \label{fig:results_lane3}
\end{figure}

\section{Data Illustration}
\label{sec:data}
There is considerable interest in evaluating the risk of heart failure linked to different diabetes treatment options \citep{hippisley2016diabetes}. To date, insulin remains on of the most effective treatment for glycemic control. Meanwhile, other medications like GLP-1 receptor agonists, DPP-4 inhibitors, and SGLT-2 inhibitors have gained prominence as alternative or adjunctive therapies to insulin. However, the impact of these treatments on long-term incidence heart failure remains unclear. Recent studies have found that non-insulin medications are associated with lower cardiovascular risk profiles \citep{wang2024comparing}, while conflicting evidence suggests the difference is not significant \citep{alkhezi2021heart}. We demonstrate the proposed methods using electronic health records from the \textit{All of Us} platform to investigate the effects of non-insulin treatments on incident heart failure compared to insulin. The \textit{All of Us} program collects health data from one million individuals and offers a diverse  platform for advancing precision medicine. While the \textit{All of Us} data is centralized, it serves as an effective case study for illustrating the performance of the federated algorithm.

\begin{figure}[htb]
    \centering
    \includegraphics[width=0.8\linewidth]{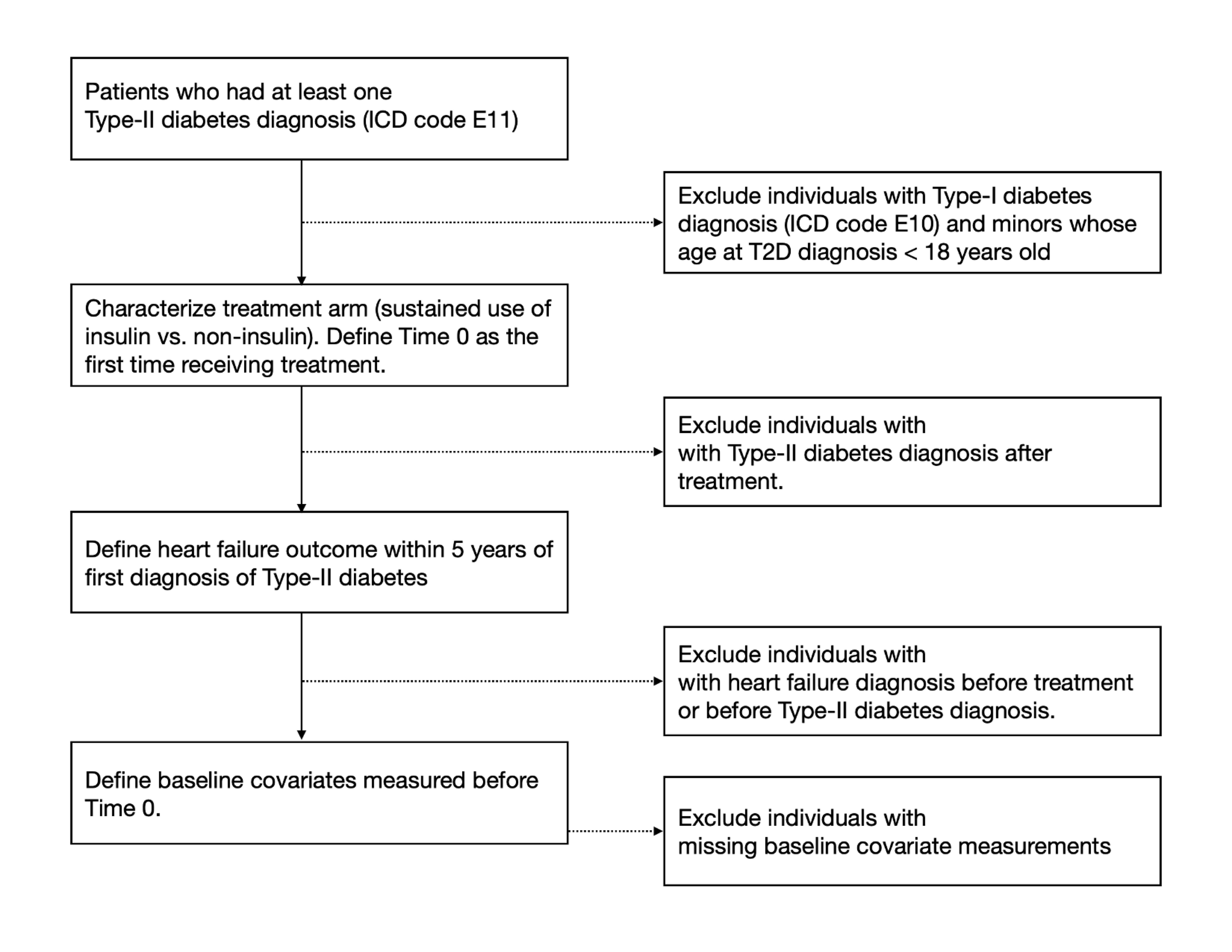}
    \caption{{Flow chart of inclusion and exclusion criteria of the study cohort}}
    \label{fig:cohort}
\end{figure}

 We define our cohort as described in Figure~\ref{fig:cohort}. We start with all patients who have at least one type 2 diabetes (T2D) billing code (ICD-10 code: E11) and define date of the T2D diagnosis as the date of the first T2D code. We exclude individuals with type I diabetes diagnosis (ICD-10 code: E10) or minors (age at T2D diagnosis less than 18 years). Next, we assign individual's treatment groups  and define the notation of ``sustained'' treatment for patients who receive multiple treatment types following \cite{wang2024comparing}. 
 We define the index date $t_0$ as the first time receiving the assigned treatment and exclude individuals whose T2D diagnosis is after $t_0$. The outcome of interest is whether one experienced a heart failure incidence within 5 years of first diagnosis of T2D, which includes congestive heart failure (ICD-10 code: I50.0), heart failure (ICD-10 code: I50), systolic or combined heart failure (ICD-9 code: 428.2) and diastolic heart failure (ICD-9 code: 428.3). We exclude patients with an observed heart failure code before $t_0$. Lastly, we adjust for the following set of baseline covariates that are measured before $t_0$ in order to eliminate unmeasured confounding: sex at birth, age at diagnosis, use of statin, use of sulfonylureas, A1C and comorbidity counts of conditions outlined in Table S4 in \cite{wang2024comparing}. We provide summary statistics in Table~\ref{tab:descriptive} and observe reasonable overlap in all covariates and treatment. Together, we have $N=733$ individuals in the treatment group (non-insulin recipients), and $N=1522$ individuals in the placebo group (insulin recipients).

Although we have pooled individual-level data, we treat the data as collected from different data centers based on patient geographic locations. Specifically, we include observations from seven states {where the prevalence of either treatment groups exceeds 20\%}: Alabama, Florida, Massachusetts, Michigan, New York, Pennsylvania, and Wisconsin. We treat each of these states as the target site, and augment the target state with the rest of source states to illustrate our methods. In real world, this translates to a practical challenge encountered when implementing federated learning across different states. Firstly, data regulations and policies vary across states, posing a significant hurdle in aggregating healthcare data for federated learning purposes. In addition, states can be considered as proxies for measuring healthcare quality, reflecting variations in medical practices, resources, and patient demographics. Consequently, state serves as an important effect modifier, influencing the outcomes of healthcare interventions. 
We assume the density ratio between the conditional density of heart failure takes the following form:
\begin{align*}
    \frac{P^0_{\textnormal{target state}}(Y\mid A,\bX)}{P^0_{\textnormal{source state}}(Y\mid A,\bX)} = \frac{\exp(\beta^0_1X_1Y + \ldots + \beta^0_6X_6Y + \beta^0_7AY)}{E_{\textnormal{target state}}[\exp(\beta^0_1X_1Y + \ldots + \beta^0_6X_6Y + \beta^0_7AY) \mid A, \bX]}. 
\end{align*}

\begin{figure}[H]
    \centering
    \includegraphics[width=0.8\linewidth]{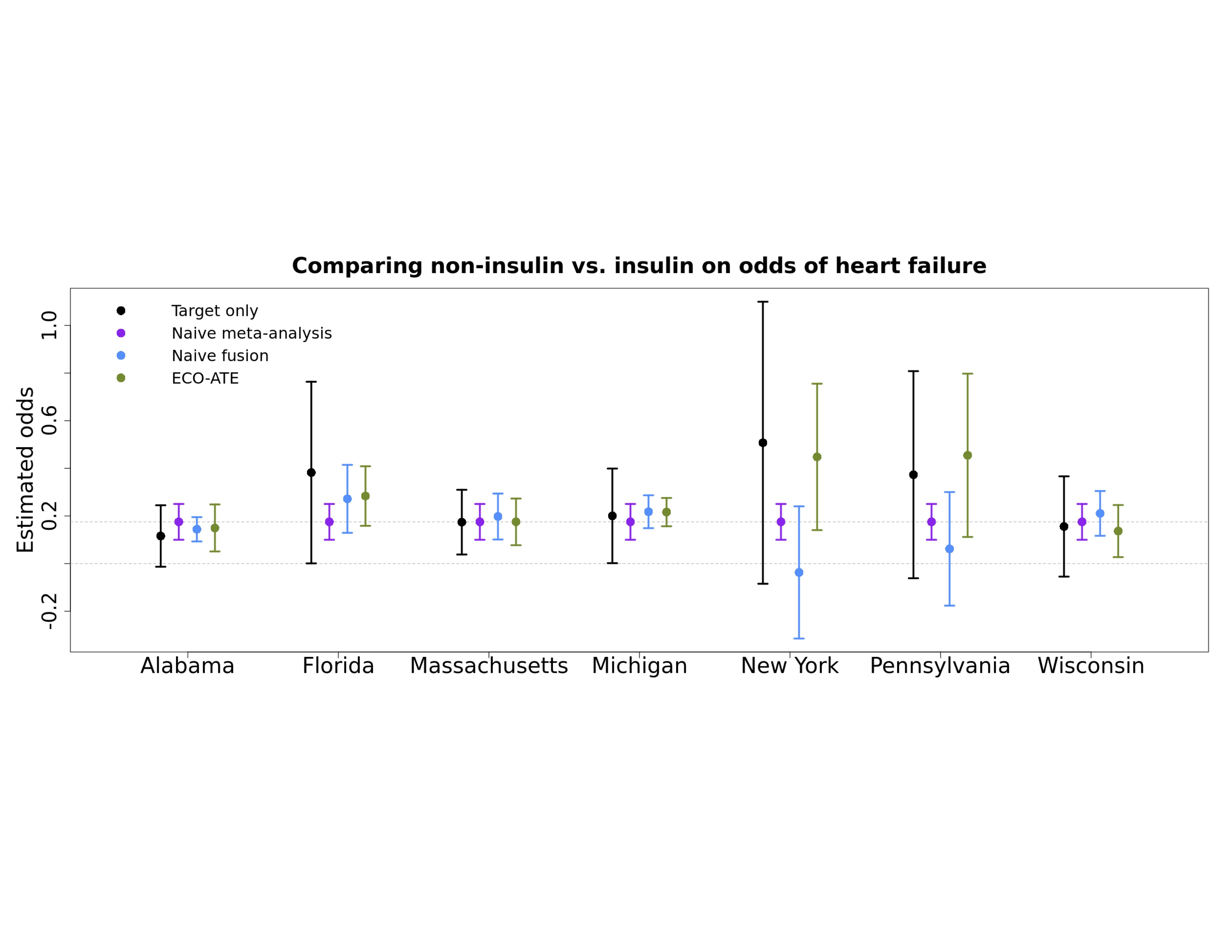}
    \caption{Estimated odds ratio of heart failure and 95\% confidence interval comparing non-insulin to insulin for patients with type II diabetes by target-only estimator (black), na\"ive meta analysis inverse variance weighting estimator (purple and grey dashed line), na\"ive fusion estimator (blue) and ECO-ATE estimator (green). The lower 95\% confidence intervals are not truncated at 0 for better visual comparison.}
    \label{fig:state}
\end{figure}

We aim to estimate the target average treatment effect of non-insulin treatments on the scale of odds ratios, and compare the following estimators: (1) the target-site only estimator, (2) the meta-analysis estimator constructed via inverse variance weighting, (3) {a na\"ive fusion estimator that assumes exchangeability in conditional outcome distributions across states}, and (4) the proposed ECO-ATE estimator. We use exponential tilt density ratio models 
for estimating shifts in covariates and treatment mechanisms. 
We estimate $\beta^0$ using kernel regressions and Super Learner, with a library consisting of generalized additive models, random forest, neural net and LASSO. Results are shown in Figure~\ref{fig:state}, with detailed numbers provided in Table~\ref{tab:state_numbers}. The target-only estimators suggest that the estimated odds of experiencing heart failure for non-insulin takers vary across states, with New York being the highest (0.507, 95\% CI [-0.085, 1.100]) and Alabama being the lowest (0.116, 95\% CI [-0.013, 0.245]). Although New York and Pennsylvania have relatively large sample size, the imbalance in treatment groups renders the resulting target-only estimator wide confidence intervals compared to other states. The na\"ive meta analysis estimator is a weighted average of all states via inverse variance weighting, and hence can only provide accurate estimate for states with state-specific odds close to the average. Similarly, the na\"ive fusion estimator assumes exchangeability in conditional outcome distributions and therefore exhibits large bias for states at the tails, i.e. Florida, New York and Pennsylvania. 
ECO-ATE reduces the variance substantially, ranging from 38\% to 91\%. For all states, our analysis suggests that non-insulin treatment leads to a lower odds of experiencing heart failure for type II diabetes patients, which is consistent with existing findings \citep{wang2024comparing}. This case study demonstrates the practical utility of the ECO-ATE algorithm in estimating causal effects of treatments within a flexibly defined target population. By relaxing the exchangeability assumption, ECO-ATE proves to be effective in accounting for site-level heterogeneity. However, we recognize that, due to the observational nature of electronic health record data and the potential for misspecification in the density ratio model, it is essential to validate these findings further. This can be achieved through goodness-of-fit tests \citep{gilbert2004goodness}, or randomized trials.

\section*{Acknowledgment}
We thank Stephanie Armbruster and Chongliang Luo for their helpful discussions. We gratefully acknowledge All of Us participants for their contributions. This work was supported by National Institutes of Health (R01 GM148494) and Patient-Centered Outcomes Research Institute (PCORI) Project Program Award (ME-2024C1-37351).\vspace*{-8pt}

\section*{Data Availability}
The raw electronic health record data used in the type-II diabetes data analysis is available on the \textit{All of Us} platform for registered researchers on the Researcher Workbench.

\section*{Appendix}

\begin{appendices}
\section{Proof to Theorem~\ref{thm:canonical}}
\label{sec:app:theorem_canonical}
We prove Theorem~\ref{thm:canonical} by first characterizing the tangent space of $\mathcal{P}$ at $P^0$. Following notations and results introduced by \cite{li2025data}, we assume that $\beta^0$ belongs to a collection $\mathcal{B}\equiv \mathbbm{R}^t$ of vectors of length $t$. For ease of notation, we define a mapping $B : \mathcal{P} \rightarrow \mathcal{B}$ such that $\beta^0\equiv B(P^0)$. 
To this end, we define $\mathcal{P}_{Q^0,\mathcal{B}}  = \{P_{Q^0,\beta}: \beta \in \mathcal{B}\}$ and $\mathcal{P}_{\mathcal{Q},\beta^0} = \{P_{Q,\beta^0}: Q\in \mathcal{Q}\}$. The former corresponds to the model where $Q^0$ is known, whereas the latter corresponds to the model where $\beta^0$ is known. We let $\mathcal{T}(P^0, \mathcal{P}_{Q^0,\mathcal{\beta}})$ and $\mathcal{T}(P^0, \mathcal{P}_{\mathcal{Q},\beta^0})$ denote the tangent space of $\mathcal{P}_{Q^0,\mathcal{B}}$ and $\mathcal{P}_{\mathcal{Q},\beta^0}$ respectively at $P^0$,  Then the tangent space of $\mathcal{P}$ at $P^0$ writes as
\begin{align*}
    \mathcal{T}(P^0,\mathcal{P}) \equiv \{ g+h: g\in \mathcal{T}(P^0, \mathcal{P}_{\mathcal{Q},\beta^0}), h\in \mathcal{T}(P^0, \mathcal{P}_{Q^0,\mathcal{\beta}})\}.
\end{align*}
In the following proof, we first derive a valid gradient for the target average treatment effect relative to $\mathcal{P}$ using the target data only and denote it as $D^\mathcal{A}_{P^0}$, following results of \cite{li2023efficient}. Then we project $D^\mathcal{A}_{P^0}$ onto the tangent space $\mathcal{T}(P^0,\mathcal{P})$ to derive the canonical gradient of the target average treatment effect $\phi(P^0)$. To begin with, it can be readily verified that a valid gradient for the target average treatment effect that uses target data only is, 
\begin{align*}
    D^\mathcal{A}_{P^0}(\bz,s) & = \frac{\mathbbm{1}(s=0)}{P^0(S=0)}\left\{\sum_{a'=0}^1 \frac{\mathbbm{1}(a=2a'-1)}{P^0(A=2a'-1 \mid \bx, S=0)} \left(y - \mu(2a'-1,\bx) \right)\right\}\\
    & + \frac{\mathbbm{1}(s=0)}{P^0(S=0)}  \left\{ \mu(1,\bx) - \mu(0,\bx) - \phi(P^0)\right\}\\
    & \equiv D^\mathcal{A}_{P^0_y}(\bz,s)+ D^\mathcal{A}_{P^0_\bx}(\bz,s).
\end{align*}
Next we project $D^\mathcal{A}_{P^0}$ onto the tangent space $ \mathcal{T}(P^0,\mathcal{P})$. In what follows, we use $\Pi_{P^0}\{ \cdot \mid \mathcal{C}\}$ to denote the $L^2_0(P^0)$-projection operator onto a subspace $\mathcal{C}$ of $L^2_0(P^0)$. Before proceeding, we let $\mathcal{T}_{\mathcal{B}}(P^0)$ denote the closed linear space spanned by the efficient score functions of $\beta^0$. We note that although $\mathcal{T}(P^0, \mathcal{P}_{\mathcal{Q},\beta^0})$ and $\mathcal{T}(P^0, \mathcal{P}_{Q^0,\mathcal{\beta}})$ are not orthogonal to each other; $\mathcal{T}(P^0, \mathcal{P}_{\mathcal{Q},\beta^0})$ and $\mathcal{T}_{\mathcal{B}}(P^0)$ are. Moreover, $ \mathcal{T}(P^0,\mathcal{P})$ equals the orthogonal sum of $\mathcal{T}(P^0, \mathcal{P}_{\mathcal{Q},\beta^0})$ and $\mathcal{T}_{\mathcal{B}}(P^0)$. Therefore,
\begin{align*}
    \Pi_{P^0}\{ D^\mathcal{A} \mid  \mathcal{T}(P^0,\mathcal{P})\}(\bz,s) & = \Pi_{P^0}\{ D^\mathcal{A} \mid  \mathcal{T}(P^0,\mathcal{P}_{\mathcal{Q},\beta^0})\}(\bz,s)  + \Pi_{P^0}\{ D^\mathcal{A} \mid  \mathcal{T}_{\mathcal{B}}(P^0)\}(\bz,s)  \\
    & = \Pi_{P^0}\{ D^\mathcal{A} \mid  \mathcal{T}(P^0,\mathcal{P}_{\mathcal{Q},\beta^0})\}(\bz,s) + E_{P^0}[D^\mathcal{A}_{P^0} \dot{\ell}^*]^\top D^\beta_{P^0}(\bz,s),
\end{align*}
where $\dot{\ell}^*$ is the efficient score function of $\beta^0$, and $D^\beta_{P^0}\equiv E_{P^0}[\dot{\ell}^*{\dot{\ell}^*}^\top]^{-1} \dot{\ell}^*(\bz,s;\beta^0)$ is the canonical gradient of $\beta^0$. By Lemma 2 in \cite{li2025data}, we have $\Pi_{P^0}\{ D^\mathcal{A} \mid  \mathcal{T}(P^0,\mathcal{P}_{\mathcal{Q},\beta^0})\}(\bz,s) =  \mathcal{L}(d)(\bz;\beta^0,P^0) - E_{P^0}[\mathcal{L}(d)(Z;\beta^0,P^0) \mid a,\bx,s]+ \frac{\mathds{1}(s =0)}{P^0(S=0)}\left(\mu_{P^0}(1,\bx)  -\mu_{P^0}(0,\bx) - \phi(P^0)\right)$.

In addition, note that 
\begin{align*}
    E_{P^0}[D^\mathcal{A}_{P^0}(\bZ,S) \dot{\ell}^*(\bZ,S)] & = E_{P^0}[\{D^\mathcal{A}_{P^0_y}(\bZ,S)+ D^\mathcal{A}_{P^0_\bx}(\bZ,S)\}\dot{\ell}^*(\bZ,S)] \\
    & = E_{P^0}[D^\mathcal{A}_{P^0_y}(\bZ,S)\dot{\ell}^*(\bZ,S)]  + E_{P^0}[D^\mathcal{A}_{P^0_\bx}(\bZ,S)\dot{\ell}^*(\bZ,S)] \\ 
    & = E_{P^0}[D^\mathcal{A}_{P^0_y}(\bZ,S)\dot{\ell}^*(\bZ,S)]. 
\end{align*}
This concludes the proof.

\section{Regularity conditions}
\label{sec:app:regularity}
We study the asymptotic properties of ECO-ATE and the required conditions. Following \cite{li2025data}, we now formalize the required alignment and overlap condition that make it possible to relate the distributions of variables of interests from source sites to the ones of the target site. 

\begin{cond}\label{cond:ide} The set $[k]$ satisfies the following:
    \begin{enumerate}[label=1\alph*,ref=S1\alph*,leftmargin=*]
        \item \label{cond:alignment}\textit{(Sufficient alignment)}: for all $s\in[k]$, $p^0(y\mid a,\bx,s) = w^*_s(\bz;\beta^0_s,W_s)p^0(y\mid a,\bx,S=0)$;
        \item \label{cond:overlap}\textit{(Sufficient overlap)}: for all $s\in\mathcal{S}$, the conditional distribution $P^0(\bX,A \mid S=0)$ is absolutely continuous with respect to the conditional distribution $P^0(\bX,A \mid S=s)$. In addition, there exists a $u_s \in [1,\infty)$ such that $Q^0( u_s^{-1} \leq \lambda^\dagger(a,\bx)/w^*_s(\bz;\beta^0_s,W_s) \leq u_s) = 1$ where we let $\lambda^\dagger(a,\bx):= p^0(\bx,a \mid S=0)/p^0(\bx,a \mid S\in\mathcal{S})$ denote the density ratio of the joint distribution of covariate and treatment mechanism between the target and all sites. 
    \end{enumerate}
\end{cond}
Condition~\ref{cond:alignment} re-iterates the semiparametric density ratio model between source and target sites. Although we use the exponential titling model in Section~\ref{sec:methods} as an example for the choice of $w$, other forms are also available \citep{bickel1993efficient} . Condition~\ref{cond:overlap} requires the site-specific density ratio of the variables of interest is bounded. This condition resembles the overlapping of site participation in \cite{han2025federated}, and the positivity of participation in \cite{dahabreh2019extending}, in the sense that we need sufficient overlap in baseline variables and treatment assignments. Additionally, the outcome $Y$ needs to share the same support between $Q^0(y\mid a,\bx)$ and $P^0(y\mid a,\bx,s)$ such that the density ratios of the outcome are bounded. We now present our main theorem, the canonical gradient of the target average treatment effect. 

We now study the conditions under which the proposed ECO-ATE estimator $\hat{\phi}_{\textnormal{ECO-ATE}}$ achieves the efficiency bound. We begin by stating the general conditions in which the one-step estimator constructed via a plug in estimate $\phi(\hat{P})$ and $D^\mathrm{eff}_{\hat{P}}$ will be asymptotic linear and efficient. {Here, $\hat{P}$ denotes a general estimator of $P^0$.}

\begin{cond}\label{lem:conditions}
    Under the following regularity conditions, the one-step estimator $\hat{\phi}:= \phi(\hat{P}) + \mathds{P}_n D^\mathrm{eff}_{\hat{P}}$ is asymptotically linear, normal and efficient: 
    \begin{enumerate}[label=2\alph*.,ref=S2\alph*,leftmargin=*]
    \item \label{cond:donsker}the empirical mean of $D^\mathrm{eff}_{\hat{P}}(\bZ,S)-D^\mathrm{eff}_{P^0}(\bZ,S)$ is within $o_p(n^{-1/2})$ of the mean of this term when $(\bZ,S)\sim P^0$, and
     \item \label{cond:remainder}the remainder term $R(\hat{P},P^0):= \phi(\hat{P}) - \phi(P^0) + E_{P^0}\{D^\mathrm{eff}_{\hat{P}}(\bZ,S)\}$ is $o_p(n^{-1/2})$.
    \end{enumerate}
\end{cond}

Condition~\ref{cond:donsker} will hold under appropriate empirical process and consistency condition. Specifically, we require $D^\mathrm{eff}_{\hat{P}}$ to be $P^0$-Donsker and the $L^2(P^0)$-norm of $(\bz,s) \rightarrow D^\mathrm{eff}_{\hat{P}}(\bz,s) - D^\mathrm{eff}_{{P}}(\bz,s)$ converges to zero in probability \citep{van2000asymptotic}. We now introduce the specific conditions on the convergence rates for the nuisance parameters to meet the requirements outlined in Condition~\ref{cond:remainder} for the ECO-ATE estimator.

\begin{cond}\label{lem:rem_conditions} We define the $j$-th component of $(k+1) \times  1$ vector $\hat{M}^-(\bX,A;\hat{\beta},\hat{W},\hat{\lambda})^\top \\ \left\{\bar{w}^*(\bZ;\hat{\beta},\hat{W})r(\bz;\hat{\beta},\hat{W},\hat{\lambda}) - E_{\hat{P}}[\bar{w}^*(\bZ;\hat{\beta},\hat{W})r(\bz;\hat{\beta},\hat{W},\hat{\lambda})\mid A,\bX,S] \right\} $ as $\widehat{R}_j(\bZ;\hat{\beta},\hat{W})$, $\delta_j(a,\bx): = {P^0}(S=j \mid a,\bx)$ and $\hat{\delta}_j(a,\bx): = \hat{P}(S=j \mid a,\bx)$. We denote the $L^2(P^0)$ norm as ${\Vert \cdot \Vert}$. Under the following conditions, the remainder term is $o_p(n^{-1/2})$. 
    \begin{enumerate}[label=3\alph*.,ref=S3\alph*,leftmargin=*]
    \item \label{cond:rem1} ${\Vert \mu(a,\bX) - \hat{\mu}(a,\bX)\Vert}{\Vert \pi(a,\bX) - \hat{\pi}(a,\bX)\Vert} = o_p(n^{-1/2})$ for $a = \{0,1\}$.
    \item \label{cond:rem2} ${\Vert \mu(a,\bX) - \hat{\mu}(a,\bX)\Vert}{\Vert\hat{\lambda}^\dagger(a,\bX) - \lambda^\dagger(a,\bX)\Vert} = o_p(n^{-1/2})$ for $a = \{0,1\}$.
    \item \label{cond:rem3} $\sum_{j \in \mathcal{S}}\Vert E_{\hat{P}}\left[\lambda_y(\bZ;\hat{\beta}, \hat{W}) (\hat{\mu}(a,\bX) - Y) \mid A=a,X,S=j\right] \Vert \Vert \hat{\delta}_j(a,\bX) - E_{P^0}[\widehat{R}_j(\bZ;\hat{\beta},\hat{W}) \mid A=a,\bX] \Vert  = o_P(n^{-1/2})$ for $a = \{0,1\}$.
    \item \label{cond:rem4} $ \left\Vert P^0(S=m \mid \bZ) - r_m(\bZ;\hat{\beta},\hat{W},\hat{\lambda}) \right\Vert  \\
     \cdot \left\Vert E_{P^0}\left[\frac{\dot{w}_m(\bZ;\hat{\beta}_m)}{w_m(\bZ;\hat{\beta}_m)} \mid A=a,\bX, S=m\right] - E_{\hat{P}}\left[\frac{\dot{w}_m(\bZ;\hat{\beta}_m)}{w_m(\bZ;\hat{\beta}_m)} \mid A=a,\bX, S=m\right]\right\Vert$ for each $m \in \mathcal{S}$ and $a = \{0,1\}$.
     \item \label{cond:rem5} $\sum_{j \in \mathcal{S}}\left\Vert E_{\hat{P}}\left[r_m(\bZ;\hat{\beta},\hat{W},\hat{\lambda})\left(\frac{\dot{w}_m(\bZ;\hat{\beta}_m)}{w_m(\bZ;\hat{\beta}_m)} - E_{\hat{P}}\left[\frac{\dot{w}_m(\bZ;\hat{\beta}_m)}{w_m(\bZ;\hat{\beta}_m)} \mid A,\bX,S=m\right]\right) \mid A=a, \bX, S=j\right] \right\Vert \\
     \lVert \hat{\delta}_j(a,\bX) - E_{P^0}[\widehat{R}_j(\bZ;\hat{\beta},\hat{W}) \mid A=a,\bX] \rVert $ for each $m \in \mathcal{S}$ and $a = \{0,1\}$.
    \end{enumerate}
\end{cond}
The ECO-ATE estimator is asymptotic linear and efficient under Conditions~\ref{cond:donsker}, \ref{cond:rem1} to \ref{cond:rem5}. 
Specifically, if the conditional outcome regressions, propensity scores, density ratios of $\bX$ and $A$, normalizing functions, and conditional expectations in the nuisance parameters are all $o_p(n^{-1/4})$, then Condition~\ref{lem:rem_conditions} is achieved. 
When $\bX$ is low-dimensional, this rate can be achieved using the methods of sieves \citep{grenander1981abstract} and other data-adaptive methods \citep{chernozhukov2018double}. It will become challenging when $\bX$ is high dimensional, this is beyond the scope of this work and we leave it to future work.  

{
\begin{remarkS} \label{rmk:mom}
    The matching of moments approach for estimating $\beta^0$ will remain valid if $P^0$ and $Q^0$ can be uniquely identified by moments. We note that there exists counter-examples where two distributions share all the moments (e.g., Chapter 11 of \cite{stoyanov2014counterexamples} and Chapter 3.15 of \cite{siegel2017counterexamples}). In these cases, we refer readers to approaches such as kernel mean matching \citep{gretton2008covariate} and discriminative learning \citep{bickel2009discriminative}. However, it is not clear how these methods can be straightforwardly extended to handle federated settings, and we leave it to future work. 
\end{remarkS}
}

\begin{proof}[\textit{to Condition S\ref{lem:rem_conditions}}]
For simplicity, we focus on the case for estimating $\phi_1(P^0):=E_{P^0}[E_{P^0}[Y\mid A=1,\bX,S=0]\mid S=0]$. For ease of notation, we denote $D^1_{P^0,y}(\bz) = \frac{\mathbbm{1}(a=1)}{P^0(A=1\mid \bx, S=0)}(y - \mu(1,\bx))$. 
The remainder term writes as 
\begin{align*}
    R(\hat{P}_\mathrm{fed},P^0) & = R_1(\hat{P}_\mathrm{fed},P^0)  + R_2(\hat{P}_\mathrm{fed},P^0), 
\end{align*}
where
\begin{align*}
R_1(\hat{P}_\mathrm{fed},P^0) 
& = \phi_1(\hat{P}_\mathrm{fed}) - \phi_1(P^0) \\
& \quad+ E_{P^0}\left[ r(\bz;\hat{\beta},\hat{W},\hat{\lambda})D^1_{\hat{P}_\mathrm{fed},y}(\bZ)  - E_{\hat{P}_\theta}[r(\bz;\hat{\beta},\hat{W},\hat{\lambda})D^1_{\hat{P}_\mathrm{fed},y}(\bZ)\mid A,\bX,S ]\right]\\
& \quad  +  E_{P^0}\bigg[ E_{\hat{P}_\theta}[r(\bz;\hat{\beta},\hat{W},\hat{\lambda})D^1_{\hat{P}_\mathrm{fed},y}(\bZ)\bar{w}^*(\bZ;\hat{\beta},\hat{W})^\top \mid A,\bX,S=0] \hat{M}^{-}(\bX,A;\hat{\beta},\hat{W},\hat{\lambda})^\top \\
& \hspace{6em}\left\{\bar{w}^*(\bZ;\hat{\beta},\hat{W}) r(\bz;\hat{\beta},\hat{W},\hat{\lambda})- E_{\hat{P}_\theta}[\bar{w}^*(\bZ;\hat{\beta},\hat{W}) r(\bz;\beta,\hat{W},\hat{\lambda})\mid A,\bX,S]\right\} \bigg] \\
& \quad  + E_{P^0}\left[\frac{\mathbbm{1}(s=0)}{P(S=0)}( \hat{\mu}(1,\bX) - \phi_1(\hat{P}_\mathrm{fed}))\right],
&\intertext{And,}
R_2(\hat{P}_\mathrm{fed},P^0) & =  - E_{\hat{P}_\mathrm{fed}}\left[\tilde{D}_{\hat{P}_\mathrm{fed}}(\bZ,S;\hat{\beta})\dot{\ell}_{\hat{P}_\mathrm{fed}}(\bZ,S;\hat{\beta})\right]^\top E_{P^0}\left[ D^\beta_{\hat{P}_\mathrm{fed}}(\bZ,S;\hat{\beta})\right].
\end{align*}
For ease of presentation, we suppress the subscripts and use $\hat{P}$ for short. We let $\lambda_y(\bz;\beta^0,W)$  denote the Radon-Nikodym derivative of the conditional outcome distribution under sampling from $Q^0$ relative to the conditional distribution $Y \mid A,\bX,S\in\mathcal{S}$ under sampling from $P^0$, that is, $\lambda_y(\bz;\beta^0,W): = q^0(y\mid a,\bx)/p^0(y \mid a,\bx,S \in \mathcal{S}) = 1/\sum_{s \in\mathcal{S}}P^0(S=s)w^*_s(\bz;\beta^0)$. Then $R_1$ reduces to
\begin{align*}
& R_1(\hat{P},P^0)  = (I) + (II)\\
& = E_{P^0}\Bigg[\hat{\lambda}^\dagger(A,\bX) \frac{\mathbbm{1}(A=1)}{\hat{\pi}(1,\bX)}  E_{P^0} \left[{\lambda}_y(\bZ;\hat{\beta},\hat{W}) \left(Y - \hat{\mu}(1,\bX)\right)\mid A,\bX,S\right] \\
& \quad + \frac{\mathbbm{1}(A=1)}{{\pi}(1,\bX)}{\lambda}^\dagger(A,\bX)\left\{\hat{\mu}(1,\bX) - \mu(1,\bX)\right\} \\
& \quad - E_{\hat{P}} \left[{\lambda}_y(\bZ;\hat{\beta},\hat{W}) \left(Y - \hat{\mu}(1,\bX)\right)\mid A,\bX,S\right] \Bigg] \\
& \quad  + E_{P^0}\Bigg[ E_{\hat{P}}[r(\bZ;\hat{\beta},\hat{W},\hat{\lambda})D^1_{\hat{P},y}(\bZ)\bar{w}^*(\bZ;\hat{\beta},\hat{W})^\top \mid A,\bX,S=0] \hat{M}^-(\bX,A;\hat{\beta},\hat{W},\hat{\lambda})^\top \\
& \hspace{5em} \cdot \left\{\bar{w}^*(\bZ;\hat{\beta},\hat{W})r(\bz;\hat{\beta},\hat{W},\hat{\lambda}) - E_{\hat{P}}[\bar{w}^*(\bZ;\hat{\beta},\hat{W})r(\bz;\hat{\beta},\hat{W},\hat{\lambda})\mid A,\bX,S] \right\} \Bigg].
\end{align*}
Now we study these terms separately. The first term can be simplified to,
\begin{align*}
(I) & =   \underbrace{E_{P^0}\left[\left\{\hat{\lambda}^\dagger(A,\bX) \frac{\mathbbm{1}(A=1)}{\hat{\pi}(1,\bX)} -{\lambda}^\dagger(A,\bX) \frac{\mathbbm{1}(A=1)}{{\pi}(1,\bX)} \right\} \left( \mu(1,\bX) - \hat{\mu}(1,\bX)\right) \right]}_{(i)}\\
& \quad + \underbrace{E_{P^0}\left[\hat{\lambda}^\dagger(A,\bX) \frac{\mathbbm{1}(A=1)}{\hat{\pi}(1,\bX)} (\hat{\mu}(1,\bX) - {\mu}(1,\bX)) \right]}_{(ii)}.
\end{align*}

By Cauchy-Schwartz Inequality, (i) can be bounded (up to a multiplicative constant) by
\begin{align*}
     \Vert \mu(1,\bX) -\hat{\mu}(1,\bX)\Vert \Vert \hat{\lambda}^\dagger(A,\bX) - \lambda^\dagger(A,\bX)\Vert + \Vert \mu(1,\bX) -\hat{\mu}(1,\bX)\Vert \Vert \hat{\pi}(1,\bX) - \pi(1,\bX)\Vert . 
\end{align*}
Note we can further write (ii) as
\begin{align*}
    (ii)& =   E_{P^0}\left[\hat{\lambda}^\dagger(A,\bX) \frac{\mathbbm{1}(A=1)}{\hat{\pi}(1,\bX)} E_{P^0}[\lambda_y(\bZ;{\beta^0},{W})(\hat{\mu}(1,\bX) - Y) \mid A,\bX \right].
\end{align*} 
\sloppy Next, we study (ii) + (II):
\begin{align*}
    & (ii) + (II) \\
    &= E_{P^0}\left[\hat{\lambda}^\dagger(A,\bX) \frac{\mathbbm{1}(A=1)}{\hat{\pi}(1,\bX)} E_{P^0}[\lambda_y(\bZ;{\beta^0},{W})(\hat{\mu}(1,\bX) - Y) \mid A,\bX \right] \\
    & \quad + E_{P^0}\left[\hat{\lambda}^\dagger(A,\bX) \frac{\mathbbm{1}(A=1)}{\hat{\pi}(1,\bX)} \sum_{j \in \mathcal{S}}E_{\hat{P}}[\lambda_y(\bZ;\hat{\beta},\hat{W})(Y - \hat{\mu}(1,\bX) ) \mid A,\bX, S=j ] \widehat{R}_j(\bZ;\hat{\beta},\hat{W})\right]\\
    &= E_{P^0}\left[\hat{\lambda}^\dagger(A,\bX) \frac{\mathbbm{1}(A=1)}{\hat{\pi}(1,\bX)}\sum_{j \in \mathcal{S}} E_{P^0}[\lambda_y(\bZ;{\beta^0},{W})(\hat{\mu}(1,\bX) - \mu(1,\bX)) \mid A,\bX, S=j] \delta_j(A,\bX) \right] \\
    & \quad + E_{P^0}\left[\hat{\lambda}^\dagger(A,\bX) \frac{\mathbbm{1}(A=1)}{\hat{\pi}(1,\bX)} \sum_{j \in \mathcal{S}}E_{\hat{P}}[\lambda_y(\bZ;\hat{\beta},\hat{W})(Y - \hat{\mu}(1,\bX) ) \mid A,\bX, S=j ] \widehat{R}_j(\bZ;\hat{\beta},\hat{W})\right]\\
    & =  E_{P^0}\Bigg[\hat{\lambda}^\dagger(A,\bX) \frac{\mathbbm{1}(A=1)}{\hat{\pi}(1,\bX)}\cdot \sum_{j \in \mathcal{S}}\\
    & \quad  \bigg\{ E_{P^0}[\lambda_y(\bZ;{\beta^0},{W})(\hat{\mu}(1,\bX) - \mu(1,\bX)) \mid A,\bX, S=j] \delta_j(A,\bX)  \\
    & \quad +   E_{\hat{P}}[\lambda_y(\bZ;\hat{\beta},\hat{W})(\mu(1,\bX) -  \hat{\mu}(1,\bX) ) \mid A,\bX, S=j] \hat{\delta}_j(A,\bX) \\
    & \quad +  E_{\hat{P}}[\lambda_y(\bZ;\hat{\beta},\hat{W})(\hat{\mu}(1,\bX)  -Y) \mid A,\bX, S=j] \left\{\hat{\delta}_j(A,\bX) - \widehat{R}_j(\bZ;\hat{\beta},\hat{W})\right\} \Bigg]\\
    & =  E_{P^0}\Bigg[\hat{\lambda}^\dagger(A,\bX) \frac{\mathbbm{1}(A=1)}{\hat{\pi}(1,\bX)} \\
    &\quad \cdot \sum_{j \in \mathcal{S}} E_{\hat{P}}[\lambda_y(\bZ;\hat{\beta},\hat{W})(\hat{\mu}(1,\bX)  -Y) \mid A,\bX, S=j] \left\{\hat{\delta}_j(A,\bX) - E_{P^0} \left[\widehat{R}_j(\bZ;\hat{\beta},\hat{W}) \mid A,\bX\right]\right\}  \Bigg].
\end{align*}

By Cauchy-Schwartz Inequality, (ii) + (II) can be bounded (up to a multiplicative constant) by
\begin{align*}
     \sum_{j \in \mathcal{S}}\Vert E_{\hat{P}}\left[\lambda_y(\bZ;\hat{\beta}, \hat{W}) (\hat{\mu}(1,\bX) - Y) \mid A,X,S=j\right] \Vert \Vert \hat{\delta}_j(A,\bX) - E_{P^0}[\widehat{R}_j(\bZ;\hat{\beta},\hat{W}) \mid A,\bX] \Vert . 
\end{align*}
Hence under Conditions~\ref{cond:rem1} to \ref{cond:rem3}, we have $(I)+ (II) = (i) + (ii) + (II)  = o_P(n^{-1/2})$.

Next, we study the second component of the remainder term associated with estimating $\beta^0$,
\begin{align*}
R_2(\hat{P},P^0) & =  - E_{\hat{P}}\left[\tilde{D}_{\hat{P}}(\bZ,S;\hat{\beta})\dot{\ell}_{\hat{P}}(\bZ,S;\hat{\beta})\right]^\top E_{P^0}\left[ D^\beta_{\hat{P}}(\bZ,S;\hat{\beta})\right].
\end{align*}
Provided that $E_{\hat{P}}\left[\tilde{D}_{\hat{P}}(\bZ,S;\hat{\beta})\dot{\ell}_{\beta}(\bZ,S;\hat{\beta},\hat{P})\right]$ is bounded, and $\hat{I}$ is invertible, it suffices to study the term $E_{P^0}\left[\dot{\ell}^*_{\beta}(\bZ,S;\hat{\beta},\hat{P})\right]$. For clarity, we study the the efficient score function for a specific $\beta^0_m$, which is the corresponding parameter for measuring the shift between source site $m$ and the target. For simplicity, we assume $\beta^0_m \in\mathbbm{R}$. Then it can be verified that, 

\begin{align}
  & E_{P^0}\left[\dot{\ell}^*_{\beta_m}(\bZ,S;\hat{\beta}_m,\hat{P})\right] \nonumber\\
  &   = E_{P^0}\left[\left(\mathbbm{1}(S=m) - r_m(\bZ;\hat{\beta},\hat{W},\hat{\lambda})\right) \left(\frac{\dot{w}_m(\bZ;\hat{\beta}_m)}{w_m(\bZ;\hat{\beta}_m)} - E_{\hat{P}}\left[\frac{\dot{w}_m(\bZ;\hat{\beta}_m)}{w_m(\bZ;\hat{\beta}_m)} \mid A,\bX,S=m\right]\right)\right]\\
  & \quad  + E_{P^0}\Bigg[\sum_{j\in \mathcal{S}} (\hat{\delta}_j(A,\bX) - \hat{R}_j(\bZ;\hat{\beta},\hat{W})) \nonumber\\ 
  & \quad \cdot E_{\hat{P}}\left[r_m(\bZ;\hat{\beta},\hat{W},\hat{\lambda})\left(\frac{\dot{w}_m(\bZ;\hat{\beta}_m)}{w_m(\bZ;\hat{\beta}_m)} - E_{\hat{P}}\left[\frac{\dot{w}_m(\bZ;\hat{\beta}_m)}{w_m(\bZ;\hat{\beta}_m)} \mid A,\bX,S=m\right]\right) \mid A, \bX, S=j\right] \Bigg].
\end{align}

By Cauchy-Schwarz inequality, the term on the first line is bounded up to a multiplicative factor by
\begin{multline*}
     \left\Vert P^0(S=m \mid \bZ) - r_m(\bZ;\hat{\beta},\hat{W},\hat{\lambda}) \right\Vert  \\
     \cdot \left\Vert E_{P^0}\left[\frac{\dot{w}_m(\bZ;\hat{\beta}_m)}{w_m(\bZ;\hat{\beta}_m)} \mid A,\bX, S=m\right] - E_{\hat{P}}\left[\frac{\dot{w}_m(\bZ;\hat{\beta}_m)}{w_m(\bZ;\hat{\beta}_m)} \mid A,\bX, S=m\right]\right\Vert .
\end{multline*}
And the term on the third line is bounded up to a multiplicative factor by
\begin{multline*}
     \sum_{j \in \mathcal{S}}\left\Vert E_{\hat{P}}\left[r_m(\bZ;\hat{\beta},\hat{W},\hat{\lambda})\left(\frac{\dot{w}_m(\bZ;\hat{\beta}_m)}{w_m(\bZ;\hat{\beta}_m)} - E_{\hat{P}}\left[\frac{\dot{w}_m(\bZ;\hat{\beta}_m)}{w_m(\bZ;\hat{\beta}_m)} \mid A,\bX,S=m\right]\right) \mid A, \bX, S=j\right] \right\Vert \\
     \lVert \hat{\delta}_j(A,\bX) - E_{P^0}[\widehat{R}_j(\bZ;\hat{\beta},\hat{W}) \mid A,\bX] \rVert . 
\end{multline*}

Hence $R_2(\hat{P},P^0) = o_p(n^{-1/2})$ under Conditions~\ref{cond:rem4} to \ref{cond:rem5}.
\end{proof}

\begin{remarkS} \label{rmk:pool vs. federated}The main difference between ECO-ATE and a one-step estimator  constructed with pooled individual-level data across sites, denoted as $\hat{\phi}_{\textnormal{POOLED}}$ is described below. In a federated setting, certain components of $P^0$, such as conditional expectations listed in (a)-(c) of Section 3.3, must be estimated in ways that they can be evaluated across sites only using summary statistics. When individual-level data can be pooled, practitioners may choose  more flexible methods for estimating $P^0$. When both estimators satisfy the regularity conditions in Section~\ref{sec:app:regularity}, there is no loss in efficiency due to  data-sharing barriers. That is, both $\hat{\phi}_{\textnormal{ECO-ATE}}$ and $\hat{\phi}_{\textnormal{POOLED}}$ achieve the semiparametric efficiency bound.
\end{remarkS}
    
{ \section{ Additional simulation results}
\label{sec:app:simulation results}
\subsection{Overparameterization schemes}
\label{sec:app:overpar}
The following overparameterization scheme was implemented in simulations in Section~\ref{sec:simulation}: 
\begin{enumerate}[leftmargin = *]
    \item \sloppy Parsimonious  $\beta^0 \in \mathbbm{R}^3$: $w_1(\bz;\beta_1^0) = \exp({\beta_{1}^0} x_6\log y)$, $w_2(\bz;\beta_2^0) = \exp(\beta^0_{2}a\log y) $, and $w_3(\bz;\beta_3^0) = \exp(\beta^0_{3}x_6a\log y)$. 
    \item \sloppy Overparameterized  $\beta^0 \in \mathbbm{R}^5$: $w_1(\bz;\beta_1^0) = \exp({\beta_{1}^0} (x_6\log y,\log y)^\top)$, $w_2(\bz;\beta_2^0) = \exp(\beta^0_{2}a\log y) $, and $w_3(\bz;\beta_3^0) = \exp(\beta^0_{3}(x_6a\log y,\log y)^\top)$. 
    \item \sloppy Overparameterized  $\beta^0 \in \mathbbm{R}^7$: $w_1(\bz;\beta_1^0) = \exp({\beta_{1}^0} (x_6\log y,\log y,x_7\log y)^\top)$, $w_2(\bz;\beta_2^0) = \exp(\beta^0_{2}a\log y) $, and $w_3(\bz;\beta_3^0) = \exp(\beta^0_{3}(x_6a\log y,\log y,x_8\log y)^\top)$. 
    \item  Overparameterized  $\beta^0 \in \mathbbm{R}^9$: $w_1(\bz;\beta_1^0) = \exp({\beta_{1}^0} (x_6\log y,\log y,x_7\log y,x_8\log y)^\top)$, $w_2(\bz;\beta_2^0) = \exp(\beta^0_{2}(a\log y, \log y)^\top) $, and $w_3(\bz;\beta_3^0) = \exp(\beta^0_{3}(x_6a\log y,\log y,x_8\log y)^\top)$. 
\end{enumerate}

\subsection{Estimation of nuisance parameters}
\label{sec:app:nuisance estimation}

The estimation of $W$ occurs twice. First, during step 1(ii), the target site estimates the distribution shift for each source site using summary statistics sent from the sources. To estimate $\beta^0_s$ for each site $s$, an estimate of $W_s$ is required. At this point, the target site can freely choose the model used to estimate $W_s$, leveraging its own individual-level data. Second, after estimating the nuisance parameters—including $\hat{W}$—the target site broadcasts them to the source sites. At this stage, it is essential to use data-adaptive methods with finite-dimensional parameters, allowing the source sites to reconstruct the models without access to individual-level data from the target site. To avoid confusion, we refer the first stage as the estimation of $\beta^0$ and the second stage as the estimation of $W$ hereafter. \label{2_times_W}

 Comparing different methods for estimating $\beta^0$ and $W$, we found that several data-adaptive approaches perform similarly when the dimension of $\bX$ is low. When the dimension of $\bX$ is moderate, such as in our setup with $\bX \in \mathbb{R}^{10}$, methods of linear regression and Lasso regression for estimating $\beta^0$ gives similar results and ECO-ATE achieves efficiency gains compared to the target only estimator (grey) (Figure\ref{fig:results_lane2}) when $W$ is estimated via gradient boosting (green). Among which, using random forest for estimating $W$ is slightly biased as shifts are larger. Meanwhile, ECO-ATE estimators have similar performances when the nuisance conditional expectations are estimated via linear regression and SuperLearner (Figure\ref{fig:results_lane4}). In our simulation settings where the dimension of $\bX$ is relatively high ($\bX \in \mathbb{R}^{10}$) relative to sample size ($n_0 = 500$), we found that estimating the normalizing function $W$ with simple, smooth models (e.g., linear or mildly regularized regressions) yields a well-behaved estimating equation for $\beta^0$, so numerical root-finding converges reliably. In practice, we also observed that the solution to \eqref{MoM1} can be sensitive to the initial values. Therefore, we recommend to search over a grid of starting points and select the solution with an objective value closest to zero.

\begin{figure}
    \centering
    \includegraphics[width=\linewidth]{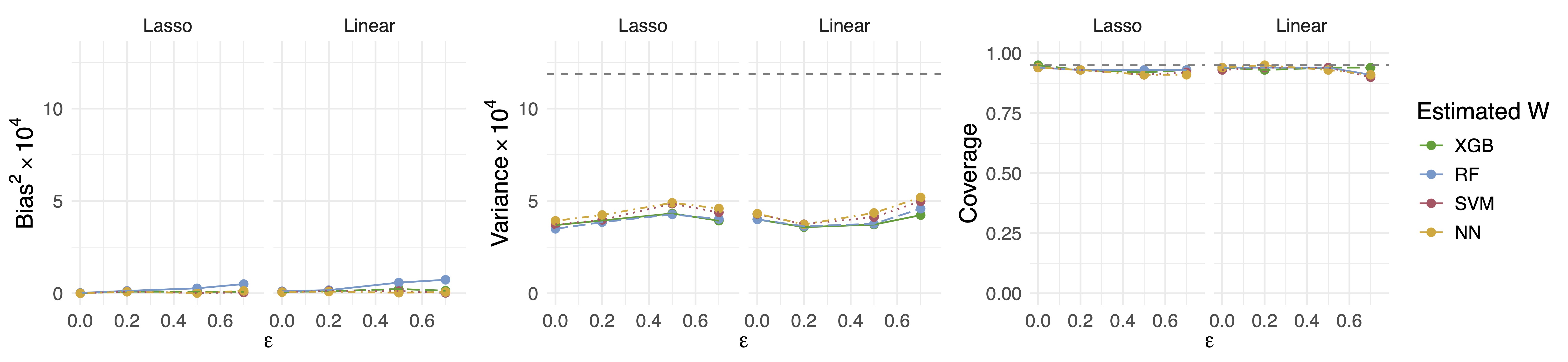}
    \caption{Scaled bias squared, variance and coverage of the proposed ECO-ATE estimators. Across panels, $\beta^0$ was estimated via linear regression and Lasso. Within each sub-figure, different models were employed to estimate $W$ as indicated by different coloring and line types.}
    \label{fig:results_lane2}
\end{figure}
\begin{figure}
    \centering
    \includegraphics[width=\linewidth]{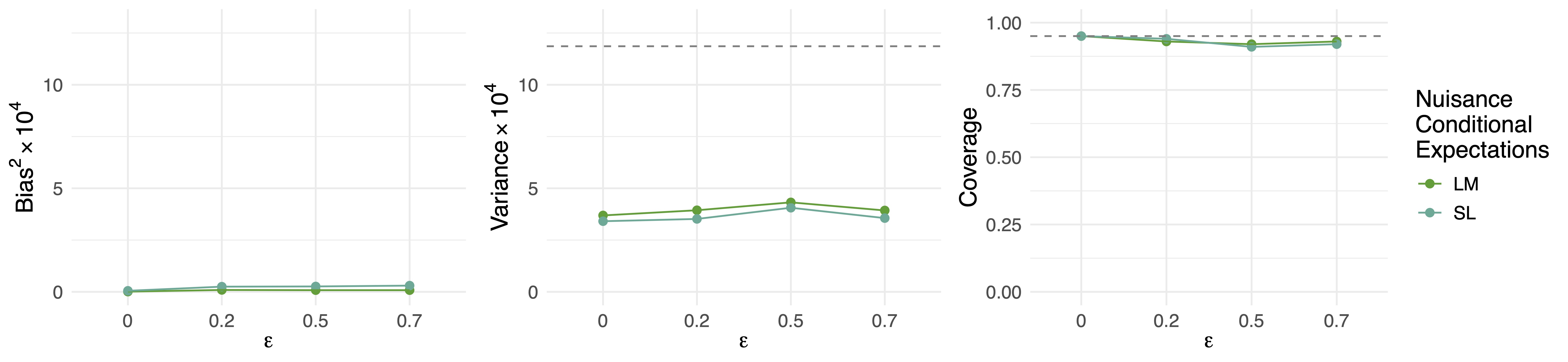}
    \caption{Scaled bias squared, variance and coverage of the proposed ECO-ATE estimators across different estimation strategies for estimating conditional expectations involving $\hat{W}$. Throughout, $\beta^0$ was estimated via LASSO and $W$ was estimated by gradient boosting.}
    \label{fig:results_lane4}
\end{figure}

\subsection{Simulation results in Tables}
\label{sec:app:sim tables}
The following tables display the bias$^2$, variance and coverage of estimators that were depicted in Figures \ref{fig:results_lane1} and \ref{fig:results_lane3}. 

\begin{table}[H]
\begin{center}
\caption{Scaled bias$^2 \times 10^4$, variance$\times 10^4$ and coverage of the target only estimator, na\"ive fusion estimator, and ECO-ATE estimators using all source sites for the target average treatment effect.  For ECO-ATE estimators, $\beta^0$ was estimated via linear regression while different models were employed for estimating $W$.   }
\label{tab:simulation_1}
\begin{adjustbox}{width=\textwidth}
\begin{tabular}{lrrrrrrrrrr}
\toprule
  & \multicolumn{3}{c}{Target only} & \multicolumn{3}{c}{Na\"ive fusion} & \multicolumn{3}{c}{ECO-ATE}\\
\cmidrule(l){2-4} \cmidrule(l){5-7}\cmidrule(l){8-10}
 & Bias$^2$ & Variance & Coverage & Bias$^2$ & Variance  & Coverage &Bias$^2$ & Variance  & Coverage\\
 \midrule
 $\epsilon = 0$\\
 \hspace{1em} XGB &0.08 & 11.72 & 0.96 & 0.04 & 3.10 & 0.96 & 0.07 & 4.00 & 0.94\\
 \hspace{1em} RF & 0.08 & 11.72 & 0.96 & 0.04 & 3.10 & 0.96 & 0.11 & 4.00 & 0.94\\
 \hspace{1em} SVM & 0.08 & 11.72 & 0.96 & 0.04 & 3.10 & 0.96 & 0.07 & 4.29 & 0.93\\
 \hspace{1em} NN & 0.08 & 11.72 & 0.96 & 0.04 & 3.10 & 0.96 & 0.06 & 4.31 & 0.94\\
$\epsilon = 0.2$\\
 \hspace{1em} XGB & 0.08 & 11.72 & 0.96 & 0.35 & 2.86 & 0.94 & 0.11 & 3.58 & 0.93\\
 \hspace{1em} RF  & 0.08 & 11.72 & 0.96 & 0.35 & 2.86 & 0.94 & 0.17 & 3.63 & 0.94\\
\hspace{1em} SVM & 0.08 & 11.72 & 0.96 & 0.35 & 2.86 & 0.94 & 0.12 & 3.74 & 0.94\\
\hspace{1em} NN  &0.08 & 11.72 & 0.96 & 0.35 & 2.86 & 0.94 & 0.09 & 3.73 & 0.95\\
$\epsilon = 0.5$\\
 \hspace{1em} XGB & 0.08 & 11.72 & 0.96 & 2.38 & 2.86 & 0.86 & 0.23 & 3.72 & 0.94\\
 \hspace{1em} RF & 0.08 & 11.72 & 0.96 & 2.38 & 2.86 & 0.86 & 0.58 & 3.76 & 0.94\\
\hspace{1em} SVM  & 0.08 & 11.72 & 0.96 & 2.38 & 2.86 & 0.86 & 0.12 & 4.12 & 0.94\\
\hspace{1em} NN & 0.08 & 11.72 & 0.96 & 2.38 & 2.86 & 0.86 & 0.03 & 4.35 & 0.93\\
$\epsilon = 0.7$\\
 \hspace{1em} XGB &0.08 & 11.72 & 0.96 & 4.23 & 2.98 & 0.80 & 0.14 & 4.23 & 0.94\\
 \hspace{1em} RF & 0.08 & 11.72 & 0.96 & 4.23 & 2.98 & 0.80 & 0.73 & 4.58 & 0.91\\
\hspace{1em} SVM  & 0.08 & 11.72 & 0.96 & 4.23 & 2.98 & 0.80 & 0.02 & 4.98 & 0.90\\
\hspace{1em} NN & 0.08 & 11.72 & 0.96 & 4.23 & 2.98 & 0.80 & 0.07 & 5.19 & 0.91\\
\bottomrule
\end{tabular}
\end{adjustbox}
\end{center}
\end{table}

\begin{table}[H]
\begin{center}
\caption{Scaled bias$^2 \times 10^4$, variance$\times 10^4$ and coverage of the target only estimator, na\"ive fusion estimator, and ECO-ATE estimators using all source sites for the target average treatment effect.  For ECO-ATE estimators, $\beta^0$ was estimated via LASSO while different models were employed for estimating $W$.   }
\label{tab:simulation_1}
\begin{adjustbox}{width=\textwidth}
\begin{tabular}{lrrrrrrrrrr}
\toprule
  & \multicolumn{3}{c}{Target only} & \multicolumn{3}{c}{Na\"ive fusion} & \multicolumn{3}{c}{ECO-ATE}\\
\cmidrule(l){2-4} \cmidrule(l){5-7}\cmidrule(l){8-10}
 & Bias$^2$ & Variance & Coverage & Bias$^2$ & Variance  & Coverage &Bias$^2$ & Variance  & Coverage\\
 \midrule
 $\epsilon = 0$\\
 \hspace{1em} XGB &0.01 & 11.76 & 0.96 & 0.00 & 2.89 & 0.95 & 0.01 & 3.69 & 0.95\\
 \hspace{1em} RF & 0.01 & 11.76 & 0.96 & 0.00 & 2.89 & 0.95 & 0.02 & 3.49 & 0.94\\
 \hspace{1em} SVM &0.01 & 11.76 & 0.96 & 0.00 & 2.89 & 0.95 & 0.00 & 3.73 & 0.94\\
 \hspace{1em} NN & 0.01 & 11.76 & 0.96 & 0.00 & 2.89 & 0.95 & 0.00 & 3.92 & 0.94\\
$\epsilon = 0.2$\\
 \hspace{1em} XGB & 0.01 & 11.76 & 0.96 & 0.41 & 2.96 & 0.92 & 0.09 & 3.94 & 0.93\\
 \hspace{1em} RF  & 0.01 & 11.76 & 0.96 & 0.41 & 2.96 & 0.92 & 0.13 & 3.85 & 0.93\\
\hspace{1em} SVM & 0.01 & 11.76 & 0.96 & 0.41 & 2.96 & 0.92 & 0.09 & 3.97 & 0.93\\
\hspace{1em} NN  &0.01 & 11.76 & 0.96 & 0.41 & 2.96 & 0.92 & 0.08 & 4.24 & 0.93\\
$\epsilon = 0.5$\\
 \hspace{1em} XGB &0.01 & 11.76 & 0.96 & 2.05 & 3.09 & 0.87 & 0.08 & 4.32 & 0.92\\
 \hspace{1em} RF & 0.01 & 11.76 & 0.96 & 2.05 & 3.09 & 0.87 & 0.27 & 4.27 & 0.93\\
\hspace{1em} SVM  & 0.01 & 11.76 & 0.96 & 2.05 & 3.09 & 0.87 & 0.02 & 4.85 & 0.91\\
\hspace{1em} NN & 0.01 & 11.76 & 0.96 & 2.05 & 3.09 & 0.87 & 0.01 & 4.90 & 0.91\\
$\epsilon = 0.7$\\
 \hspace{1em} XGB &0.01 & 11.76 & 0.96 & 4.09 & 2.92 & 0.81 & 0.08 & 3.93 & 0.93\\
 \hspace{1em} RF & 0.01 & 11.76 & 0.96 & 4.09 & 2.92 & 0.81 & 0.50 & 4.02 & 0.93\\
\hspace{1em} SVM  & 0.01 & 11.76 & 0.96 & 4.09 & 2.92 & 0.81 & 0.04 & 4.38 & 0.92\\
\hspace{1em} NN & 0.01 & 11.76 & 0.96 & 4.09 & 2.92 & 0.81 & 0.13 & 4.59 & 0.91\\
\bottomrule
\end{tabular}
\end{adjustbox}
\end{center}
\end{table}

\begin{table}[H]
\begin{center}
\caption{Scaled bias$^2 \times 10^4$, variance$\times 10^4$ and coverage of the ECO-ATE estimators for the target average treatment effect with overparameterized density ratio models. For ECO-ATE estimators, $\beta^0$ was estimated via LASSO and $W$ was estimated by gradient boosting. }
\label{tab:simulation_3}
\begin{adjustbox}{width=\textwidth}
\begin{tabular}{lrrrrrrrrrrrrr}
\toprule
  & \multicolumn{3}{c}{$\epsilon = 0$} & \multicolumn{3}{c}{$\epsilon = 0.2$} &\multicolumn{3}{c}{$\epsilon = 0.5$} & \multicolumn{3}{c}{$\epsilon = 0.7$}\\
\cmidrule(l){2-4} \cmidrule(l){5-7}\cmidrule(l){8-10}\cmidrule(l){11-13}
 & Bias$^2$ & Variance & Coverage & Bias$^2$ & Variance  & Coverage &Bias$^2$ & Variance  & Coverage& Bias$^2$ & Variance  & Coverage\\
 \midrule
$\beta^0 \in \mathbbm{R}^3$&  0.01 & 3.69 & 0.95 & 0.09 & 3.94 & 0.93 & 0.08 & 4.32 & 0.92 & 0.08 & 3.93 & 0.93\\
$\beta^0 \in \mathbbm{R}^5$&0.11 & 4.15 & 0.92 & 0.28 & 4.21 & 0.91 & 0.35 & 5.06 & 0.88 & 0.28 & 4.46 & 0.92\\
$\beta^0 \in \mathbbm{R}^7$& 0.02 & 4.13 & 0.93 & 0.16 & 4.38 & 0.91 & 0.17 & 5.23 & 0.90 & 0.49 & 4.72 & 0.90\\
$\beta^0 \in \mathbbm{R}^9$ &0.02 & 4.15 & 0.93 & 0.07 & 4.50 & 0.92 & 0.49 & 5.43 & 0.89 & 0.32 & 4.41 & 0.92\\
\bottomrule
\end{tabular}
\end{adjustbox}
\end{center}
\end{table}

}

{ \section{Extended implementation details for diabetes treatments on heart failure analysis}
\label{sec:app:data illustration algorithm}
Our illustrative example involves EHR obtained from the ``All of Us” Research Program, hypothetically partitioned into seven distinct data centers corresponding to individual states (Alabama, Florida, Massachusetts, Michigan, New York, Pennsylvania, and Wisconsin, and other states were excluded due to limited sample sizes). In reality, ``All of Us" collects health data from a diverse population of over one million participants, but for demonstration purposes, we treat each state as a standalone data center operating under strict privacy regulations. Each site’s EHR includes de-identified patient demographics (e.g., sex at birth, age at diagnosis), diagnosis codes (ICD-10 or ICD-9), laboratory values (A1C), medication exposures (insulin, SGLT-2, GLP-1, DPP-4, statin, sulfonylureas), and limited comorbidity indicators. By assigning states as separate centers, we replicate real-world data heterogeneity: diverse practice patterns, resource availability, and patient demographics, all of which often lead to systematic differences in the risk of heart failure among adults with type II diabetes. In many practical scenarios, statewide policies and institutional review board (IRB) requirements do not allow pooling individual-level data; this makes decentralized analysis essential for studying how various diabetes treatments influence HF incidence across multiple contexts.

    We now go through in detail the proposed ECO-ATE algorithm taking Alabama as the target site for example. To begin with, Alabama will estimate distribution shifts for all other six states. Specifically, each source state $s \in  $ \{Florida, Massachusetts, Michigan, New York, Pennsylvania, and Wisconsin\} sends its sample size $n_s$, estimated covariates and treatment density ratios (covariate includes sex at birth, age at diagnosis, use of statin, use of sulfonylureas, A1C and comorbidity counts), form of $w_s := \exp(\beta^0_s (\bX Y,AY)^T)$, and the corresponding summaries $\bar{\xi}_s:= \mathbbm{P}_{n,s} \xi_s(\bZ_i)$ where $\xi_s(\bZ) = (\bX Y, AY) \in \mathcal{R}^7$ to Alabama. For covariates and treatment density ratios, source site $s$ can either estimate its own $\hat{p}(\bx,a\mid S=s)$ and send the relevant model parameters to Alabama, or can send summary statistics to Alabama, who will in turn estimate the density ratios by methods such as exponential tilting models.  In Section~\ref{sec:data}, we used exponential tilting models for estimating $\lambda_s$ with basis functions $(1,\bX)$ and $(A,A\cdot \bX)$ for covariate and treatment respectively. 
    
    Next, Alabama will solve for $\beta^0_s$ for each $s$ by matching moments in \eqref{MoM1}. After obtaining all $\hat{\beta}:= (\hat{\beta}_s)_{s\in [6]}$, Alabama will broadcast the following estimated nuisance parameters to all six source sites: 
    \begin{enumerate}[label=(\alph*)]
        \item Each state's: sample size, covariate and treatment density ratios $\hat{\lambda}_s$, $\hat{\beta}_s$, form of $\xi_s$, $\hat{W}_s$, model parameters for estimating $E_{P^0}[r\bar{w}^* \mid A,\bX,S= \text{Alabama}]$ and $E_{P^0}[r\bar{w}^*{\bar{w}^*}^\top \mid A,\bX,S= \text{Alabama}]$.
        \item Estimated propensity scores and outcome regression model for Alabama: $\hat{\pi}$ and $\hat{\mu}$. Model parameters for estimating $E_{P^0}[\tilde{d} \mid A,\bX,S= \text{Alabama}]$ and $E_{P^0}[\tilde{d}\bar{w}^* \mid A,\bX,S= \text{Alabama}]$.
        \item Model parameters for estimating $E_{P^0}[\tilde{a} \mid A,\bX,S= \text{Alabama}]$ and $E_{P^0}[\tilde{a}\bar{w}^* \mid A,\bX,S= \text{Alabama}]$.
    \end{enumerate}

    In Section~\ref{sec:data}, we employed SuperLearner for estimating $W_s$ with a library consisting of generalized additive models, LASSO, random forest, and neural network. All conditional expectations were estimated via  main terms linear regression. Propensity score was estimated via main terms linear-logistic regression.

    Now, each state $s$ has received the list of estimated nuisance parameters from Alabama and can therefore construct $\mathcal{L}_s$, $\mathcal{I}_s$ and $\mathcal{H}_s$ and send them back to Alabama. Using these quantities, Alabama will construct $(\mathcal{M}_s)_{s\in \mathcal{S}}$, in addition to
    the same set of quantities $\mathcal{L}_0$, $\mathcal{I}_0$ and $\mathcal{H}_0$. Lastly, the proposed ECO-ATE estimator can be constructed as $\hat{\phi}_{\textnormal{ECO-ATE}} =  \frac{1}{7}\sum_{s\in \mathcal{S} }(\mathcal{H}_s + \mathcal{M}_s) + \mathcal{N}_0$. }

\section{Additional data illustration results}
\label{sec:app:data illustration}
The following tables contain the descriptive summaries of the study cohort, and detailed estimated odds, variance and 95\% Confidence Interval for each state. 
\begin{table}[H]
\begin{center}
\caption{Descriptive summaries of the study cohort.}
\label{tab:descriptive}
\begin{adjustbox}{width=\textwidth}
\begin{tabular}{lrrrrrrr}
\toprule
& \multicolumn{1}{c}{Sex at birth} & \multicolumn{1}{c}{Age at diagnosis} & \multicolumn{2}{c}{Baseline Medications} & \multicolumn{2}{c}{Baseline Clinical Features} & \multicolumn{1}{c}{Heart Failure}\\
\cmidrule(l){2-2} \cmidrule(l){3-3} \cmidrule(l){4-5} \cmidrule(l){6-7} \cmidrule(l){8-8}
 & Female  &  Mean (SD) & Statin & Sulfonylureas & Comor counts & A1C &  Yes\\
\hline
\textbf{Alabama}\\
\hspace{1em} Insulin (N=83)& 58 (69.9\%) & 52.3 (11.3) & 29 (34.9\%) & 32 (38.6\%) & 1.96 (1.70) & 8.14 (1.85) & 4 (4.8\%)\\
\hspace{1em} Non-insulin (N=73) & 51 (69.9\%) & 51.8 (11.0) & 11 (15.1\%) & 18 (24.7\%) & 1.15 (1.27) & 7.77 (2.36) &  17 (23.3\%)\\
\textbf{Florida }\\
\hspace{1em} Insulin (N=83) & 55 (66.3\%)& 56.8 (9.96) & 33 (39.8\%) & 26 (31.3\%) & 1.60 (1.41) & 7.75 (1.61) & 8 (9.6\%)\\
\hspace{1em} Non-insulin (N=44)  &  34 (77.3\%)  & 57.1 (12.7) & 10 (22.7\%) & 12 (27.3\%) & 1.75 (1.46) & 7.56 (1.91) & 10 (22.7\%)\\
\textbf{Massachusetts} \\
\hspace{1em} Insulin (N=398) &  221 (55.5\%)  & 55.2 (10.2) & 200 (50.3\%) & 161 (40.5\%) & 2.62 (2.18) & 8.14 (1.88) &  8 (2.0\%)\\
\hspace{1em} Non-insulin (N=262) & 129 (49.2\%)  & 55.6 (12.0) & 79 (30.2\%) & 102 (38.9\%) & 2.70 (2.23) & 8.20 (2.22) &  28 (10.7\%)\\
\textbf{Michigan} \\
\hspace{1em} Insulin (N=128) & 84 (65.6\%)  & 54.0 (12.5) & 43 (33.6\%) & 52 (40.6\%) & 1.78 (1.68) & 7.89 (1.79) & 5 (3.9\%)\\
\hspace{1em} Non-Insulin (N=97) & 63 (64.9\%) & 54.8 (11.8) & 20 (20.6\%) & 29 (29.9\%) & 1.90 (1.91) & 8.07 (2.31) &  18 (18.6\%)\\
\textbf{New York} \\
\hspace{1em} Insulin (N=291) & 189 (64.9\%)  & 56.3 (11.5) & 112 (38.5\%) & 102 (35.1\%) & 1.87 (1.99) & 8.15 (1.94) & 10 (3.4\%)\\
\hspace{1em} Non-Insulin (N=80) & 52 (65.0\%)  & 53.4 (15.0) & 10 (12.5\%) & 24 (30.0\%) & 1.53 (1.92) & 8.60 (2.75) & 7 (8.8\%)\\
\textbf{Pennsylvania} \\
\hspace{1em} Insulin (N=393) & 240 (61.1\%)  & 55.1 (10.2) & 200 (50.9\%) & 182 (46.3\%) & 2.06 (1.75) & 7.48 (1.81) &  9 (2.3\%)\\
\hspace{1em} Non-Insulin (N=105) & 73 (69.5\%)  & 52.8 (13.1) & 31 (29.5\%) & 45 (42.9\%) & 1.87 (1.87) & 7.78 (2.14) &  7 (6.7\%)\\
\textbf{Wisconsin} \\
\hspace{1em} Insulin (N=146) & 87 (59.6\%)  & 55.9 (11.8) & 70 (47.9\%) & 56 (38.4\%) & 2.42 (2.33) & 7.69 (1.60) & 3 (2.1\%)\\
\hspace{1em} Non-Insulin (N=72) & 43 (59.7\%)  & 53.3 (11.3) & 20 (27.8\%) & 37 (51.4\%) & 2.47 (2.28) & 8.58 (2.19) &  7 (9.7\%)\\
\textbf{Overall} \\
\hspace{1em} Insulin (N=1522) & 934 (61.4\%) & 55.3 (10.9) & 687 (45.1\%) & 611 (40.1\%) & 2.15 (1.98) & 7.89 (1.84) &  47 (3.1\%)\\
\hspace{1em} Non-Insulin (N=733) & 445 (60.7\%)  & 54.3 (12.4) & 181 (24.7\%) & 267 (36.4\%) & 2.11 (2.06) & 8.12 (2.29) &  94 (12.8\%)\\
\bottomrule
\end{tabular}
\end{adjustbox}
\end{center}
\end{table}

\begin{table}[H]
\begin{center}
\caption{Estimated odds ratio, variance and 95\% CI for each state. }
\label{tab:state_numbers}
\begin{adjustbox}{width=\textwidth}
\begin{tabular}{lrrrrrrrrrrrrr}
\toprule
  & \multicolumn{3}{c}{Target only} & \multicolumn{3}{c}{Na\"ive Meta} &\multicolumn{3}{c}{Na\"ive Fusion} & \multicolumn{3}{c}{Efficient Fusion}\\
\cmidrule(l){2-4} \cmidrule(l){5-7}\cmidrule(l){8-10}\cmidrule(l){11-13}
 & Est & Var & 95\% CI & Est & Var & 95\% CI & Est & Var & 95\% CI& Est & Var & 95\% CI\\
 \midrule
Alabama & 0.116 & 0.004 & (-0.013, 0.245) & 0.175 & 0.001 & (0.1, 0.25) & 0.144 & 0.001 & (0.093, 0.195) & 0.150 & 0.003 & (0.051, 0.248)\\

Florida & 0.382 & 0.038 & (0.001, 0.764) & 0.175 & 0.001 & (0.1, 0.25) & 0.272 & 0.005 & (0.129, 0.415) & 0.284 & 0.004 & (0.159, 0.409)\\

Massachusetts & 0.174 & 0.005 & (0.038, 0.310) & 0.175 & 0.001 & (0.1, 0.25) & 0.198 & 0.002 & (0.101, 0.294) & 0.175 & 0.002 & (0.077, 0.273)\\

Michigan & 0.200 & 0.010 & (0.002, 0.399) & 0.175 & 0.001 & (0.1, 0.25) & 0.218 & 0.001 & (0.149, 0.287) & 0.216 & 0.001 & (0.157, 0.276)\\

New York & 0.507 & 0.091 & (-0.0085, 1.100) & 0.175 & 0.001 & (0.1, 0.25) & -0.037 & 0.020 & (-0.315, 0.241) & 0.448 & 0.025 & (0.141, 0.755)\\

Pennsylvania & 0.373 & 0.049 & (-0.062, 0.808) & 0.175 & 0.001 & (0.1, 0.25) & 0.062 & 0.015 & (-0.177, 0.300) & 0.455 & 0.031 & (0.112, 0.798)\\

Wisconsin & 0.156 & 0.012 & (-0.055, 0.366) & 0.175 & 0.001 & (0.1, 0.25) & 0.211 & 0.002 & (0.117, 0.305) & 0.137 & 0.003 & (0.027, 0.246)\\
\bottomrule
\end{tabular}
\end{adjustbox}
\end{center}
\end{table}

\end{appendices}

\bibliography{Bib}

\end{document}